\documentclass{jfm}
\usepackage{graphicx}
\usepackage{epstopdf, epsfig}
\usepackage{amsmath}
\usepackage{float}
\usepackage{stmaryrd}

\usepackage{bm}
\usepackage{subcaption}
\usepackage{color}
\usepackage{xcolor}

\shorttitle{Guidelines for authors}
\shortauthor{M. Vernet}

\title{Turbulence in electromagnetically-driven  Keplerian flows}

\author{M. Vernet$^1$, M. Pereira$^2$, S. Fauve$^1$, C. Gissinger$^{1,3}$}

\affiliation{
\aff{1 }Laboratoire de Physique de l'Ecole Normale Superieure, CNRS, PSL Research University,
Sorbonne Universite, Universite de Paris, F-75005 Paris, France
\aff{2} Arts et Metiers Institute of Technology, CNAM, LIFSE, HESAM University, F-75013 Paris, France
\aff{3} Institut Universitaire de France (IUF)
}

\begin{document}

\maketitle

\begin{abstract}

 The flow of an electrically conducting fluid in a  thin disc under the  action of an azimuthal Lorentz force  is studied experimentally.   At small forcing, the Lorentz force is balanced by  either viscosity or inertia, yielding quasi-Keplerian velocity profiles. For very large current and moderate magnetic field, we observe a new regime, fully  turbulent, which exhibits large  fluctuations and a Keplerian mean rotation profile $\Omega\sim \frac{\sqrt{IB}}{r^{3/2}}$. In this turbulent regime, the dynamics is typical of thin layer turbulence, characterized by a direct cascade of energy towards the small scales and an inverse cascade to large scale. Finally, at very large magnetic field, this  turbulent flow bifurcates to a quasi-bidimensional turbulent flow  involving the formation of a large scale condensate in the horizontal plane. These results are well understood as resulting from an instability of the B\"odewadt-Hartmann  layers at large Reynolds number and  discussed in the  framework  of similar astrophysical flows. 
 


  \end{abstract}


\section{Background}

 ---------------------------
 
 Since the pioneering work of Hartmann \citep{Hartmann37} on electromagnetic conduction pumps, several works have been devoted to the study of  flows of liquid metals driven by a stationary electromagnetic force. Experimentally, it is usually easier to study this problem in a cylindrical geometry, in which the flow is driven in the azimuthal direction under the effect of an axial magnetic field and the injection of a radial current. Because of the relatively homogeneous and delocalized forcing, this configuration is also advantageous over other setups when it comes to generate turbulent rotating flows. 
 
 In 1965, \citet{Hunt65} proposed one of the first theoretical studies of a magnetohydrodynamic flow in the presence of conductive walls and deduced a theoretical expression for electrically-driven flows. A first experimental confirmation of this theory was obtained in 1971 by \citet{Bayliss71} and \citet{Bayliss71b}, in a laboratory experiment using mercury. By imposing a magnetic field strong enough to inhibit inertial effects, the authors confirmed the linear dependence of the flow with the applied current $I$.
Later,  \citet{Tabeling81} proposed a new theoretical approach by looking at the recirculation likely to occur when the external field  is not too strong. They found that a secondary flow  involving no inertial effects is produced by Hartmann's boundary layers. On the contrary, Shercliff layers (boundary layers parallel to the field) produce recirculation
allowing inertial effects to penetrate the heart of the flow, thus modifying the average velocity profile when the Reynolds number is sufficient large.
Similarly, \citet{Potherat00} analytically predicted two typical phenomena likely to occur in such electromagnetically driven flows in annular geometry: first, a secondary recirculation related to inertial effects in the Hartmann boundary layer, and second the possibility of  a transverse dependence of the velocity in the bulk flow. These predictions were confirmed experimentally by the same authors. It is now well accepted that the main action of the magnetic field is thus to reduce the velocity gradients in its direction, and that the nature of the transition between strongly magnetized bidimensional flows and 3D flows is intimately linked to recirculations \citep{Sommeria82, Potherat12}. Due to the substantial number of control parameters, there are still several unanswered questions regarding the different flow regimes observed in electrically-driven annular duct flow.

More recently, several studies have focused on the destabilization of such flows for moderate Reynolds numbers, when the Hartmann layers become unstable. The experimental study by \citet{Moresco04} looked at the distribution of currents in the flow and reported how the destabilization of Hartmann's boundary layers is controlled by the parameter
$Re_H=Re/Ha$ (Table \ref{table:dimnum}) which is a critical Reynolds number based on the thickness of the laminar Hartmann layer. In this experience, the transition to turbulence is observed for values of $Re_H$ two orders of magnitude smaller than what is predicted by linear stability theory, thus suggesting the existence of a subcritical transition in this type of flow. These results have been completed by direct numerical simulations \citep{Krasnov04}, proposing an explanation for this difference between simulations and experiments based on the growth of disturbances of finite amplitude near the boundary layers. In some cases it has been shown that the instability of the flow is rather controlled by the Reynolds number based on the thickness of the Shercliff layer \citep{Potherat07}.
Moreover, it has also been shown numerically that current-driven flows in annular geometry can also become unstable at moderate Hartmann numbers, via centrifugal-type instability \citep{Zhao11,Zhao12}

 Such destabilization is also reported in other laboratory experiments. In \citet{Tabeling81}, it was shown that electrically-driven flows can undergo a sequence of instabilities involving slow oscillations  just before the transition to turbulence. In tall cylindrical apparatus, \citet{Boisson12,Boisson17} have shown that  traveling waves can appear when the forcing is large enough due to inertial terms. More recently, a  Kelvin-Helmoltz destabilisation of free shear layers generated near the electrodes has also been observed \citep{Stelzer15,Stelzer15b}. Finally, note that \citet{Messadek02} also found that at very large Reynolds and Hartmann numbers, in a configuration involving a sheared layer, the flow exhibits quasi-two-dimensional turbulence. \\
 
 Most of the laboratory experiments described above rely on the generation of large forcing by the use of very strong magnetic field. For example, the experiences of \citet{Messadek02}  reach $IB\sim 300 AT$ in order to obtain high Reynolds number regimes. However, this is achieved by using magnetic fields relatively large (between $1$ and $10$ Tesla). Thus, these laboratory experiments describe inertial regimes, but always in a strongly magnetized limit, for which the interaction parameter $N_t$ (Table \ref{table:dimnum}) is large. One of the motivation of the experiment reported here is to go  beyond the classical inertial regime and to reach a fully turbulent state at small $N_t$. As shown in the next sections, this is made possible by three important features of the KEPLER experiment: a particularly thin disc geometry (compared to the usual square section), a driving which mostly relies on a very large current (rather than strong magnetic field) and a large horizontal size leading to large Reynolds numbers.\\

 Finally, the study of the fully turbulent flow of a magnetohydrodynamic thin disc finds direct application in astrophysics. For example, accretion disks, one of the most studied problems in astrophysical fluid dynamics, involve turbulent flows in which a fluid is in Keplerian rotation around a massive central body, usually a star, a protostar or a black hole. The exact mechanisms by which the angular momentum is transported outward in these discs remain unknown. Indeed, the enormous accretion rates observed by astrophysicists indicate that the transport of material inward, losing angular momentum, must be compensated by a large transport of angular momentum outward. Keplerian rotation profiles being hydrodynamically stable according to the Rayleigh criterion for centrifugal instability, a hydrodynamic linear instability cannot be invoked to explain this angular momentum transport \citep{Rayleigh17,Ji06,Velikhov06,Fromang19}. In order to understand the existence of this regime, different mechanisms have been proposed like subcritical transition to turbulence \citep{Lesur05}, extraction of angular momentum by MHD winds from the disk \citep{Lesur21}, or the role of density stratification \citep{Dubrulle05}. But the most accepted scenario is the magnetorotational instability (MRI) \citep{Balbus91}, which explains how a conducting fluid in differential rotation subjected to a sufficiently large magnetic field can be destabilized towards a fully turbulent state. Although extensively studied numerically and theoretically, the experimental observation of MRI remains a major challenge for
modern fluid dynamics \citep{Sisan04,Hantao12,Stefani06}, mainly due to the difficulty in achieving a high magnetic Reynolds number in the laboratory, but also because of the difficulty to generate a stable Keplerian flow. In this regard, it has been proposed that electrically-conducting boundaries \citep{Winarto20} and electromagnetically-driven flows might be an efficient configuration for stabilizing Rayleigh-unstable flows near the boundaries \citep{Stefani04} and satisfy the required conditions for observing MRI instability in the laboratory \citep{Khalzov10}. The interest of our experiment with respect to this problem is therefore twofold: first, it reproduces the thin-disc configuration, the flow configuration and the presence of a magnetic field typical of astrophysical disks, but it also elucidates the regime of electromagnetically-driven flows at large Reynolds number.


\section{Experimental setup}\label{sec:rules_submission}
\subsection{General  description}

\begin{figure}
    \centering
    \includegraphics[scale=0.5]{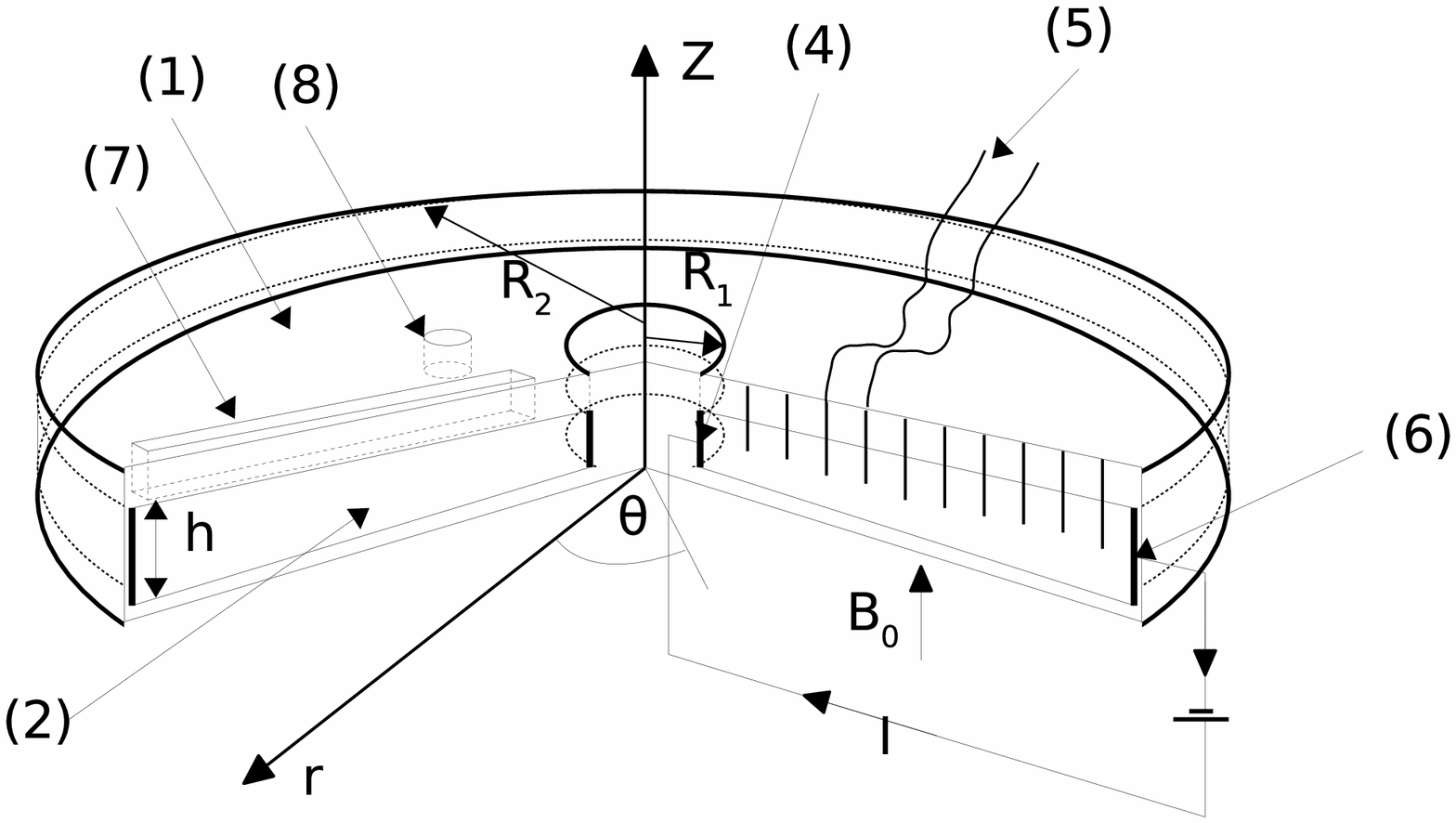}
    \caption{Sketch of the experimental setup. A thin disc of diameter $D=38$cm and thickness $h=1.5$cm is filled with a liquid metal (Galinstan) and placed between two Helmholtz coils generating a vertical homogeneous magnetic field. The cell is  confined by top and bottom walls (1) made of electrically insulating plexiglass  and  by electrically conducting cylindrical walls acting as an anode (4) and cathode (6). The cross section of the duct is rectangular (2), and characterized by a large aspect ratio $\Delta r/h\simeq 9$. The flow is entirely driven by the Lorentz force generated by the combined action of the vertical magnetic field and the strong radial current passing from the anode to the cathode.Velocity measurements are made by potential probes inserted in the top wall up to the middle of the gap (5) and thanks to Doppler probes inserted in a rail inclined at an angle of $45^{\circ}$ analogous to (7) not represented here for clarity. Induced magnetic field measurements were done thanks to Hall probes inserted in a rail (7) and the pressure sensor was inserted in the top wall (8).}
    \label{fig:kepler}
\end{figure}

The KEPLER experiment (see Fig.\ref{fig:kepler}) consists of an annular cylindrical channel with an internal  diameter  $2R_i=12$cm and an external cylinder diameter $2R_o=38$cm (mean radius $r_m=12.5$cm, gap $\Delta r=13$cm). It is filled with \textit{Galinstan}, an eutectic alloy of Gallium, Indium and Tin, liquid at room temperature whose physical characteristics are summarized in Table \ref{table:gal}.  A strong electrical current (up to $3000$ Ampere) is injected radially by a direct current generator \textit{Power TEN P66 Series 53000} from the inner cylinder to the outer rim. The temperature of the inner cylinder is controlled by a water cooling system. Both electrodes are made of brass and protected from the Galinstan by an electrochemical deposit of Nickel, in order to ensure a good electric contact with the liquid metal. The height of the cell is $h=1.5$cm, corresponding to a relatively large aspect ratio $\Gamma=\Delta r/h= 8.7$.
The top and bottom walls are electrically insulating  plexiglas plates. The cylinder is placed between two large Helmholtz coils generating a homogeneous vertical magnetic field along the vertical $z$-axis. The coils are powered by two direct current generator \textit{Ametek RS 20V-250A}, producing magnetic fields up to $110$mT. The combination of the radial current and the vertical magnetic field generates a strong Lorentz force in the azimuthal direction.

\begin{table}
    \centering
    \begin{tabular}{cc}
    \hline
    Name & Galinstan MCP11 \\
    \hline
    Chemical composition & $68.5\%$ Ga, $21.5\%$ In, $10\% Sn$ \\
    \hline
    Density & $6.44\cdot 10^3 kg.m^{-3} (293K)$ \\
    \hline
    Electric conductivity & $3.46\cdot 10^6 S.m^{-1}$ \\
    \hline
    Melting point & $254K$ \\
    \hline
    Kinematic viscosity & $3.7\cdot 10^{-7} m^2.s^{-1}$ \\
    \hline 
    Sound speed & $2730 m.s^{-1}$ \\
    \hline
    \end{tabular}
    \caption{}
    \label{table:gal}
\end{table}

\subsection{Measurement methods}
\label{sub:measure}

The velocity field is recorded using two different measurement techniques. First, velocity profiles are measured with Ultrasonic Doppler Velocimetry (UDV), by using the
commercial system \textit{DOP3010} from \textit{Signal Processing SA},  in which probes send ultrasonic pulses at regular intervals. The phase shift between the sent pulse and its echo on particles in the bulk (mostly oxydes  in liquid metals) is computed in order to obtain velocity profile along the direction pointed by  the Doppler probe. The Doppler probes are installed into the plexiglass plates at an angle of $45^{\circ}$ with the flow. Plexiglass and Galinstan have a close acoustic impedance, ensuring a good transmission of the ultra-sonic pulses.  Note that due to the angle, the measured velocity is a linear combination of the local azimuthal component of the velocity $u_{\theta}$ and the vertical component $u_z$. 

In addition,  both azimuthal and radial components of the velocity are  also measured using potential probes. 
For a conductive fluid in motion, Ohm's law specifies that the current density $\bm{j} = \bm{\nabla}\times\bm{B}/\mu_0$ is given by the relation $\bm{j} = -\sigma \bm{\nabla}\phi + \sigma (\bm{u}\times \bm{B})$. After having removed the injected current, the local velocity  is deduced from the potential difference between two probes separated by a small distance $d$ using  the relation $U_{\theta} = \delta\phi/(B_0 d)$ where $B_0$ is the magnitude of the vertical magnetic field.
Nine small electrodes ($1$mm diameter, separated from each other by $d=8$mm) are inserted equidistantly into the top plate through holes located between inner and outer cylinders. These probes are Nickel wires electrically insulated everywhere but at the tip. The mean component of the velocity is obtained through measurement of the voltage by a nanovoltmeter \textit{Keithley Model 182} and  filtered via a filter-amplifier \textit{Stanford} Research Systems Model SR560 in DC mode. The turbulent fluctuations are  obtained  from amplification of the signal through a filter-amplifier \textit{Princeton Applied Research Low Noise Amplifier Model 1900}  using smaller potential probes ($d=4$mm).

Pressure fluctuations are measured by a pressure sensor \textit{Kistler Type701A} (sensibility of $82.24pC/bar$, acquisition frequency $10$kHz), placed in contact with the liquid metal through a hole in the top plexiglass plate and coupled to a charge  amplifier \textit{Kistler Type5018}.
Finally,  induced magnetic field were measured  by a Teslameter \textit{F.W. Bell Model7030} connected  to Hall  probes (precision up to $\sim 10\mu T$) inserted in vertical holes in the plexiglass plate around two millimeters away from the liquid metal.

\section{ Length scales and dimensionless numbers of the experiment}

The dynamics of an incompressible electrically conducting fluid is  described by the MHD equations, which couple the Navier-Stokes equation to the induction equation. The latter is obtained from a combination of Maxwell's equations and the Ohm's law, such that the system of equations is:

\begin{align}
    \rho\frac{\partial\bm{u}}{\partial t} + \rho\bm{u}\cdot\nabla\bm{u} = -\nabla p + \rho\nu\Delta\bm{u} + \bm{j}\times\bm{B}\\
    \frac{\partial B}{\partial t} = \nabla\times (\bm{u}\times\bm{B}) + \eta\Delta\bm{B}\\
    \nabla\cdot\bm{u}=0 ~~ \nabla\cdot\bm{B}=0
\end{align}

where $\bm{u}$ is the velocity field, $\bm{B}$ the magnetic field, $\bm{j}$ the current density, $\rho$ the liquid metal density, $\nu$ is the kinematic viscosity and $\eta=1/(\mu_0\sigma)$ is the magnetic diffusivity (where $\mu_0$ is the magnetic permeability and $\sigma$ is the electrical conductivity). \\

In addition to the physical properties of the fluid  ($\rho$, $\nu$, $\sigma$, $\mu_0$) and the geometrical parameters of the experiment ($h$, $r_m$, $\Delta r$), the typical azimuthal velocity $U$ of the fluid is controlled by the applied current $I_0$ and magnetic field $B_0$. Our experiment is therefore described by a set of six independent dimensionless numbers, the two geometrical numbers $\Gamma=\Delta r/h$, $h/r_m$, and the  kinetic Reynolds number $Re$ and $Ha$

\begin{align}
    \textit{Re} = \frac{Ur_m}{\nu} ~~ \text{and}~~ \textit{Ha} = B_0 h \left(\frac{\sigma}{\rho\nu}\right)^{\frac{1}{2}}
\end{align}

The Reynolds number compares advection to diffusion and the Hartmann number is a dimensionless measure of the Lorentz force (compared to viscous dissipation). Note that an important characteristic of the {\it KEPLER} experiment  is the use of very high current $I_0\sim 3000$Amps, which leads to large Reynolds number ($Re> 10^{5}$)  while the Hartmann number remains moderate ($Ha<60$). In KEPLER experiment, $Ha\sim 0.57B_0$ with $B_0$ in $mT$. As discussed later, when the flow becomes quasi-bidimensional, a better definition of the Reynolds number is $Re_h=Uh/\nu$.

One can also define a magnetic Reynolds number $Rm=Ur_m/\eta$, which compares induction to magnetic diffusion in the induction equation. Alternatively, the magnetic Prandtl number $Pm=\nu/\eta$ can be used instead of $Rm$. In liquid metals, the kinematic viscosity is much smaller than the magnetic resistivity ($Pm=Rm/Re< 10^{-5}$). It follows that the characteristic time of the fluctuations of the magnetic field $\tau_b \sim l^2/\eta$ is far smaller than $\tau_{\nu}\sim l^2/\nu$ (i.e $\tau_b/\tau_{\nu} \ll 1$).   Therefore, in the limit of small induction ($Rm\ll1$), $\bm{b}$  follows $\bm{u}$ in a "quasi-static" way and the induction equation becomes \citep{Moffatt}:

\begin{align}
    0 \simeq \bm{B_0}\cdot\nabla \bm{u} + \eta\Delta\bm{b}
\end{align}
Because our experiment operates hardly outside this regime ($Rm=1$ at most),  MHD phenomena requiring large induction effects are not expected. This rules out the possibility of observing  dynamo action or MRI  instability for instance.

The dimensionless current is defined as :

\begin{align}
    \Upsilon=\frac{I_0}{2\pi r_mh\sigma UB_0}
\end{align}

 It is the ratio between the current  density  $j_0=\frac{I_0}{2\pi rh}$ directly injected between the radial electrodes and the current $j_{ind}\simeq\sigma u_\theta B_0$ induced  in the radial direction by  the  motion of the fluid (at  lowest order).
 Since this induced current is mostly generated in the bulk flow, it controls  the distribution of the  current: at low $\Upsilon$, most of the current passes through the top and bottom boundary layers, while it occupies the whole height of the cell at $\Upsilon\gg1$ .\\

In principle, these six dimensionless numbers are sufficient to fully describe the flow. However it is useful to define two additional dimensionless ratios. The first one is related to the boundary layers generated at the vertical walls of the experiment: two different types of boundary layers can be generated  in such MHD rotating flows at the top and bottom boundaries. First, Hartmann layers arise when a shear flow approaches walls perpendicular  to a magnetic field. These boundary layers  are characterized  by a  force  balance  between the Lorentz force and the viscous dissipation inside a typical thickness $\delta_{\textit{Ha}}$ given by $\delta_{\textit{Ha}}^{-1}\sim B\sqrt{\sigma/\rho\nu}$. These layers are known\citep{Murgatroyd53,Moresco04} to become unstable when the Reynolds number of the boundary layer $Re_H=\frac{U\delta_{\textit{Ha}}}{\nu} = \frac{\textit{Re}}{\textit{Ha}}$ becomes sufficiently large.
On the other hand, purely hydrodynamical boundary layers can also be generated at the Hartmann walls: as the rapidly rotating fluid reaches the boundaries where viscous forces dominate, an imbalance between the decreasing centrifugal force and the radial pressure gradient produces a radial flow in a boundary layer of typical thickness $\delta_{\Omega} = (\nu/\Omega)^{1/2}$.
Such B\"odewadt layers can also exhibit an instability at large Reynolds number. 
Following \citet{Davidson02}, we introduce the Elsasser number defined by:

\begin{align}
   \Lambda = \frac{\sigma B_0^2}{\rho \Omega}=\frac{\delta_{\Omega}^2}{\delta_{\textit{Ha}}^2}
\end{align}

As a ratio between the Lorentz and centrifugal forces, the Elsasser number tells which effect dominates the establishment of the boundary layer close to the walls.

%
%
%
%

One last important dimensionless number is the interaction parameter (also known as the Stuart number) which is the ratio between the Lorentz force and inertia, up to some geometric factor.
\begin{align}
    N_t =  \frac{\sigma B_0^2 h}{\rho U}\left(\frac{\Delta r}{h}\right)^2 = \frac{\textit{Ha}^2}{\textit{Re}}\left(\frac{\Delta r}{h}\right)^3
\end{align}

$N_t$ quantifies the strength of the magnetic field compared to the inertia of the flow. It has been shown that this {\it true} interaction parameter accurately describes the dimensionality of MHD flows \citep{Sommeria82,Messadek02, Potherat14}. As $N_t$ becomes large, the velocity field becomes independent of the direction in which the magnetic field is applied. In particular, if both $Re$ and $N_t$ are  sufficiently large, a quasi-bidimensional turbulent flow is expected.

\begin{table}
    \centering
    \begin{tabular}{ccc}
    \hline
    $Re$ & $Ur_m/\nu$ & $>10^5$ \\
    \hline
    $Re_h$ & $Uh/\nu$ & $>10^4$\\
    \hline
    $Ha$ & $B_0h(\sigma/\rho\nu)^{\frac{1}{2}}$ & $<60$ ($\sim 0.57B_0[mT]$) \\
    \hline
     $Rm$ & $Ur_m/\eta$ & $\ll 1$ \\
    \hline
    $Pm$ & $Rm/Re = \nu/\eta$ & $\ll 1$ \\
    \hline
    $\Upsilon$ & $I_0/2\pi r_mh\sigma U B_0$ & $j_0/j_{ind}$ \\
    \hline 
    $Re_H$ & $U\delta_{Ha}/\nu$ & $Re/Ha$ \\
    \hline 
    $Re_B$ & $U\delta_\Omega/\nu$ & $ Re^{1/2}$ \\
    \hline
    $\Lambda$ & $\sigma B_0^2/\rho\Omega$ & $\delta_{\Omega}^2/\delta_{Ha}^2$ \\
    \hline
    $N_t$ & $\sigma B_0^2h/\rho U (\Delta r/h)^2$ & $Ha^2/Re(\Delta r/h)^3 ~ \sim 10\Lambda$  \\
    \hline
    $\delta_{\Omega}$ & $\sqrt{\nu/\Omega}$ & $\sim r_m/Re^{1/2}$  \\
    \hline
    $\delta_{Ha}$ &  $\sqrt{\rho\nu/\sigma B_0^2}$ & $\sim h/Ha$ \\
    \hline 
    \end{tabular}
    \caption{}
    \label{table:dimnum}
\end{table}

%
%
%

\begin{figure}
    \centering
    \includegraphics[scale=0.3]{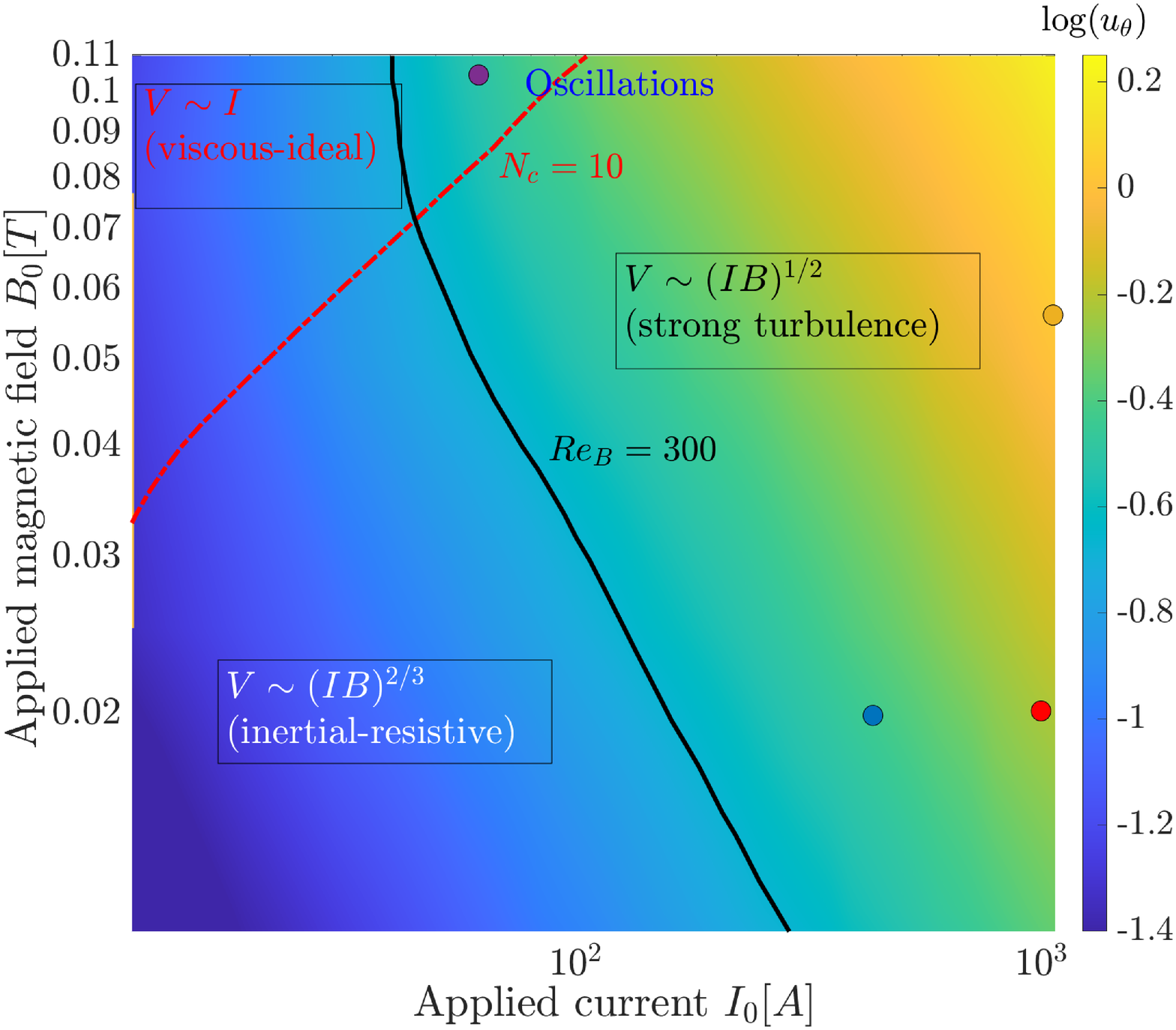}
    \caption{Map of the logarithm of the mean azimuthal flow, $\log(\overline{u_\theta})$ as a function of $(I_0,B_0)$, measured with the potential probes. The red dashed line ($N_c=10$) and the black solid line ($Re_B=300$) delimitate the transition between the different regimes generated in the experiment. The points indicate the four typical runs discussed in section 6.}
    \label{fig:map_patch}
\end{figure}

\subsection{Parameter space and outline of the results}
Fig. \ref{fig:map_patch} summarizes the various results obtained with the KEPLER experiment. The colorplot shows interpolated values of $log(u_\theta)$ in the ($I_0,B_0$) parameter space. Four different domains have been identified. These regimes will be described in more details in the next sections, but it is insightful to give a short overview of the parameter space now: at large magnetic field ($N_t>10$, indicated by the red dashed line) and small Reynolds number ($Re<9000$, indicated by the black solid line ), the {\it viscous-ideal} regime corresponds to a balance between the Lorentz force and the viscous dissipation, and is characterized by a strong value of the external magnetic field such that the induction term $\bm{u}\times\bm{B}_0$ dominates the magnetic diffusion term $\eta\Delta\bm{b}$ where $\bm{b}$ is the induced field. The flow is laminar, and the velocity field scales as  $u_\theta\propto I_0/r$, independently of $B$. 

At smaller magnetic field but moderate Reynolds number, the {\it inertial-resistive} regime is obtained. The Lorentz force is now balanced by inertia, but the boundary layers remain laminar. In addition, a significant amount of current passes through the bulk flow. The corresponding velocity field scales as $u_\theta\propto (IB)^{2/3}r^{-1/3}f(\Delta r/r)$ ($f$ being an unknown function) and remains in a quasi-Keplerian state.
Note that the first regime has been identified in previous MHD experiments and  well described theoretically \citep{Bayliss71, Messadek02}. Although not discussed in details by previous experimental studies, analysis of published data suggest that the second regime has also been observed in previous experiments.

However, when the Reynolds number is increased, the flow then bifurcates to a {\it quasi-bidimensional turbulent flow} regime, in which both the bulk and the boundary layers are turbulent. In this so-called {\it ultimate regime}, we observe a new scaling $u_\theta\propto \sqrt{IB/r}$, corresponding to a turbulent Keplerian velocity profile sharing some similarities with the velocity profile expected in accretion discs. This turbulent regime is observed only if the magnetic field is not too large, for $N_t<10$, and therefore requires an extremely large applied current.

Finally, at large magnetic field ($N_t>10$) and large Reynolds number ($Re>9000$), we observe a bi-dimensional turbulent regime which exhibits slow dynamics and condensation of the energy at large scale.

\section{Flow regimes and transition to turbulence}

Most of previous experimental studies focused either on the laminar regime (obtained at small $Re$, large $Ha$) or on the regime of quasi-bidimensional turbulence (rather observed at large $N_t$, large $Re$). The novelty of the KEPLER experiment  lies in the use of very large applied current (maximum current $I_0^m=3000$ A) and a moderate applied magnetic field (maximum field $B_0^m\sim0.11$T). The forcing, controlled by the product $I_0B_0$, can therefore reach very large values ($I_0^mB_0^m\sim 300$ TA) while keeping the Hartmann number small ($Ha=60$ at most). By comparison, \citet{Messadek02} reach comparable values, $I_0^mB_0^m\sim 500$ TA, but only at very large magnetic fields ($B_0\sim 5T$), corresponding to $Ha\sim 1500$. Here, the smallness of $Ha$ makes possible to observe a full destabilization of the Hartmann layers, leading to flows significantly more turbulent. On the other hand, the relatively weak values of our applied magnetic fields are somehow compensated by the large aspect ratio of the experiment $\Delta r/h\sim 9$ and the large value of $\Delta r$. This leads to values of the interaction parameter $N_t$ not too small, allowing to observe quasi-bidimensional turbulence in some parameter range.

To see how these different regimes are generated, Fig. \ref{fig:vIB}(a) shows time-averaged measurements obtained from the potential probes located at a radial distance $r=12$cm, approximately in the middle of the cylindrical gap. We plot the mean value of $u_\theta$ as a function of the applied current $I_0$, for different values of the magnetic field, ranging from $B_0=6$mT to the highest possible value $B_0=110$mT. The typical velocities of the flow span a large range of values, from 1 cm.s$^{-1}$ to a few meter per second, yielding $Re\sim 10^{5}$ at most. Since the applied field can both drive flow motions and laminarize it, Fig.\ref{fig:vIB} shows the same data but with $u_\theta$ plotted as a function of $I_0B_0$ instead. Three different scaling laws can be deduced from these data.

\subsection{ Viscous-ideal flows }

\begin{figure}
    \centering
    \includegraphics[scale=0.25]{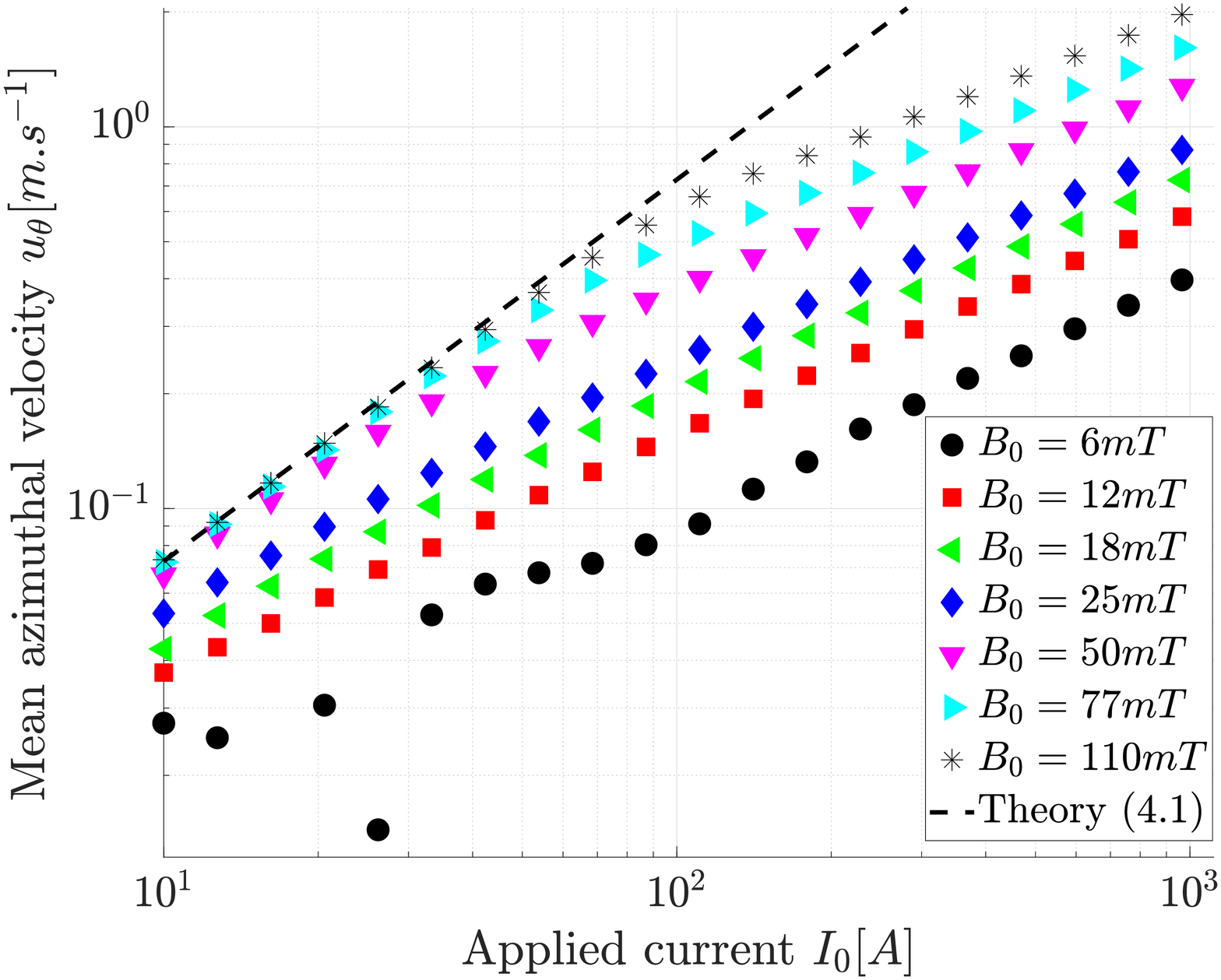}
    \includegraphics[scale=0.27]{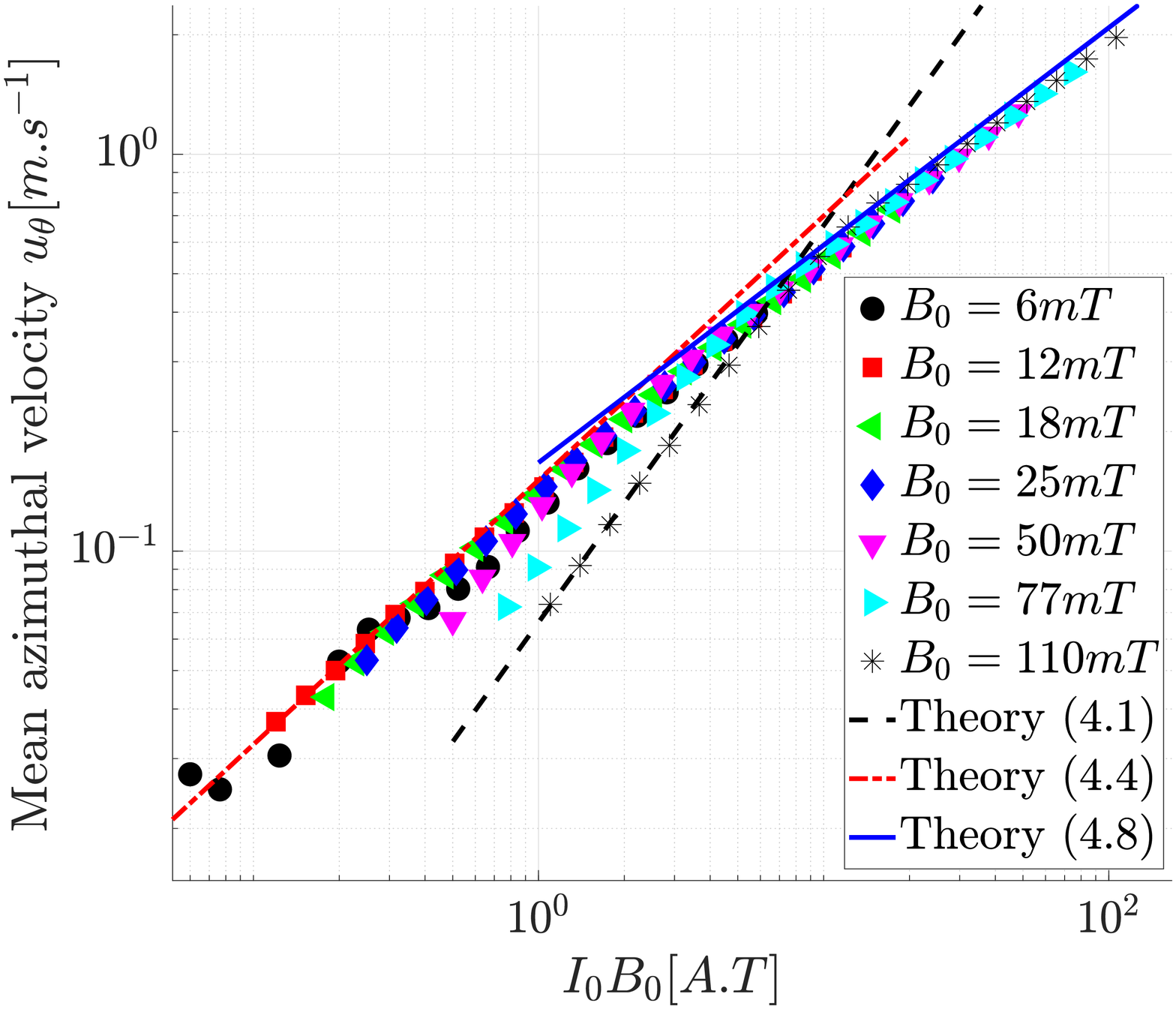}

    \caption{( a) Mean azimuthal flow $\overline{u_{\theta}}$ versus the applied current product $I_0$ measured by the potential probes for different values of the applied field $B_0$. The velocity is measured at a radial distance from the center $r=12cm$. (b) Same, plotted as a function of the product $I_0B_0$. Three different scaling laws can be observed, indicated by the solid line ($u_\theta\propto \sqrt{I_0B_0}$), the dashed line ($u_\theta\propto I_0$) and the dash-dotted line ($u_\theta\propto (I_0B_0)^{2/3}$). $Ha\sim 0.57B_0$ with $B_0$ in $mT$.}
    \label{fig:vIB}
\end{figure}

 At large magnetic field  ($Ha\ge30$), Fig.\ref{fig:vIB}(a) shows that the velocity scales linearly with the applied current at low $I_0$. 
This regime has been first described theoretically by  \citet{Bayliss71b} for a laminar axisymmetric flow in a toroidal channel of square cross section, and later confirmed by most experimental studies mentioned in the introduction.  

The theoretical prediction for this laminar flow can be retrieved by simple arguments. When the Elsasser number $\Lambda$ is sufficiently large, the Hartmann layers are much thinner than B\"odewadt layers, meaning that most of the velocity gradient occurs in layers of size $\delta_{Ha}$. In addition, if $\Upsilon$ is small, most of the applied current $I_0$ passes through the two Hartmann layers with a current density ${\bf j}\simeq \frac{I_0}{2\pi r (2\delta_{Ha})}{\bf e_r}$ in each layer.  The azimuthal Lorentz force $F_\theta=j_rB_0$ is then balanced in the Hartmann  layer by the viscous dissipation $\rho\nu \Delta\bm{u}\sim \rho\nu u_\theta/\delta_{\textit{Ha}}^2\bm{e_{\theta}}$, leading to:

\begin{equation}
U_\theta = \frac{I_0}{4\pi r\sqrt{\rho\nu\sigma}}
\label{viscous_ideal}
\end{equation}

%
%
%
%

In this expression, $B_0$ does not appear explicitly, although it controls the bi-dimensionalization of the flow. Note that this expression stands only if the Shercliff boundary layers generated at the surface of the cylinders parallel to the magnetic field do not interfere with the bulk flow. This means $r_2-r_1\gg \frac{h}{\sqrt{\textit{Ha}}}$, which is well satisfied in the KEPLER experiment. Note that the  {\it ideal} term used here refers to the fact that magnetic diffusion term $\eta\Delta\bm{b}$ can be neglected compared to the induction $uB_0$. It is however important to mention that this comes from the fact that  the field $B_0$ is strong, and is quite different from the classical {\it ideal MHD} generally obtained when $Rm\gg 1$.

The velocity field obtained at large $B_0$ and low current agrees very well with this prediction, indicated by the dashed line in  Fig.\ref{fig:vIB}.

\subsection{ Inertial-resistive flows }
Fig.\ref{fig:vIB}(b) shows that in fact, two different scaling laws can be obtained at low forcing: for $Ha\ge30$, the linear scaling $u_\theta\propto I_0$ of the {\it viscous-ideal} regime described above, and a scaling of the form $u_\theta \propto (I_0B_0)^{2/3}$ at smaller $Ha$. A similar regime seems to have been observed in at least one previous experiments \citep{Bayliss71} and one recent numerical study \citep{Poye20}. As shown by Fig.\ref{fig:vIB}, this regime is observed for $I_0B_0<10$ and $B_0<40$mT. We give below simple arguments to explain this regime, following closely \citet{Poye20}. 
 
 In the bulk flow,  inertia terms in the azimuthal direction are balanced by the applied Lorentz force $\frac{I_0B_0}{2\pi rh}$. In contrast to the previous case, experimental values of $\Upsilon$ suggest that the fraction of current passing through the bulk should be relatively large. It follows that:
 \begin{equation}
u_\theta \simeq \frac{I_0B_0}{2\pi\rho h u_r}f\left(\frac{\Delta r}{r}\right)
\label{u23_1}
\end{equation}

where the inertial term $\rho u_ru_\theta/r$ is larger than $\rho u_r\partial_ru_\theta$ because $r$ is smaller than the gap $\Delta r$ in the inner part of the disc. At large radii, $r>\Delta r$ and the second term might be dominant, which is taken into account by the function $f$. Because $r\sim \Delta r$ in any cases, the function $f$ is expected to be of order one and the choice of the inertial term will not change the scaling law.

For moderate magnetic fields, when $\Lambda\leq1$,  the radial inflow near the boundaries is mostly dominated by the B\"odewadt layers. The corresponding imbalance between the centrifugal force in the layer and the pressure gradient $\partial_rp\sim\rho u_\theta^2$ leads to 

\begin{equation}
u_r^{BL}\sim \frac{\delta_B^2u_\theta^2}{\nu r}
\label{u23_2}
\end{equation}
where $u_r^{BL}$ is the typical radial velocity in the B\"odewadt layer of thickness $\delta_B$. Because $u_z$ is presumably small in the bulk flow, incompressibility implies $\int_0^hu_rdz\sim 0$ at lowest order and so $u_rh\sim 2\delta_Bu_r^{BL}$. By combining this last result, with expressions (\ref{u23_1}) and (\ref{u23_2}), one finally obtains:

\begin{equation}
u_\theta = \left(\frac{I_0B_0}{4\pi\rho\sqrt{\nu}} \right)^{\frac{2}{3}}r^{-\frac{1}{3}}f(\frac{\Delta r}{r})
\label{inertial_resistive}
\end{equation}

Note that the unknown function $f$ may influence the radial dependance, but not the scaling law for the magnitude of the flow because $f\sim 1$. Although the scaling law (\ref{inertial_resistive}) shares some similarities with the one found by \citet{Poye20}, our full expression is quite different, and leads to values one order of magnitude smaller. This is due to some differences between our experiment and the setup studied by these authors, in particular the relatively large aspect ratio used here ($\Delta r/h\sim 9$) and the fact that $r_m\sim\Delta r$ in the KEPLER experiment. When plotting the results of \citet{Bayliss71} in this form, it appears that their data follow a similar dependance $u_\theta\propto(I_0B_0)^{2/3}$. A dependance $u_\theta\propto(I_0)^{2/3}$ based on a different argument was also observed by \citet{Potherat14}, although their setup is quite different. At the exception of  \cite{Messadek02} where a very turbulent flow is reached, we argue that this regime is probably the one mostly observed in previous experiments operating at large Reynolds number.


Prediction \ref{inertial_resistive}  (with $f=1$) is shown by the dash-dotted line in Fig.\ref{fig:vIB} . It fits our experimental data with a very good agreement.

%
%

\subsection{Keplerian turbulence}

\begin{figure}
    \centering
    \includegraphics[scale=0.25]{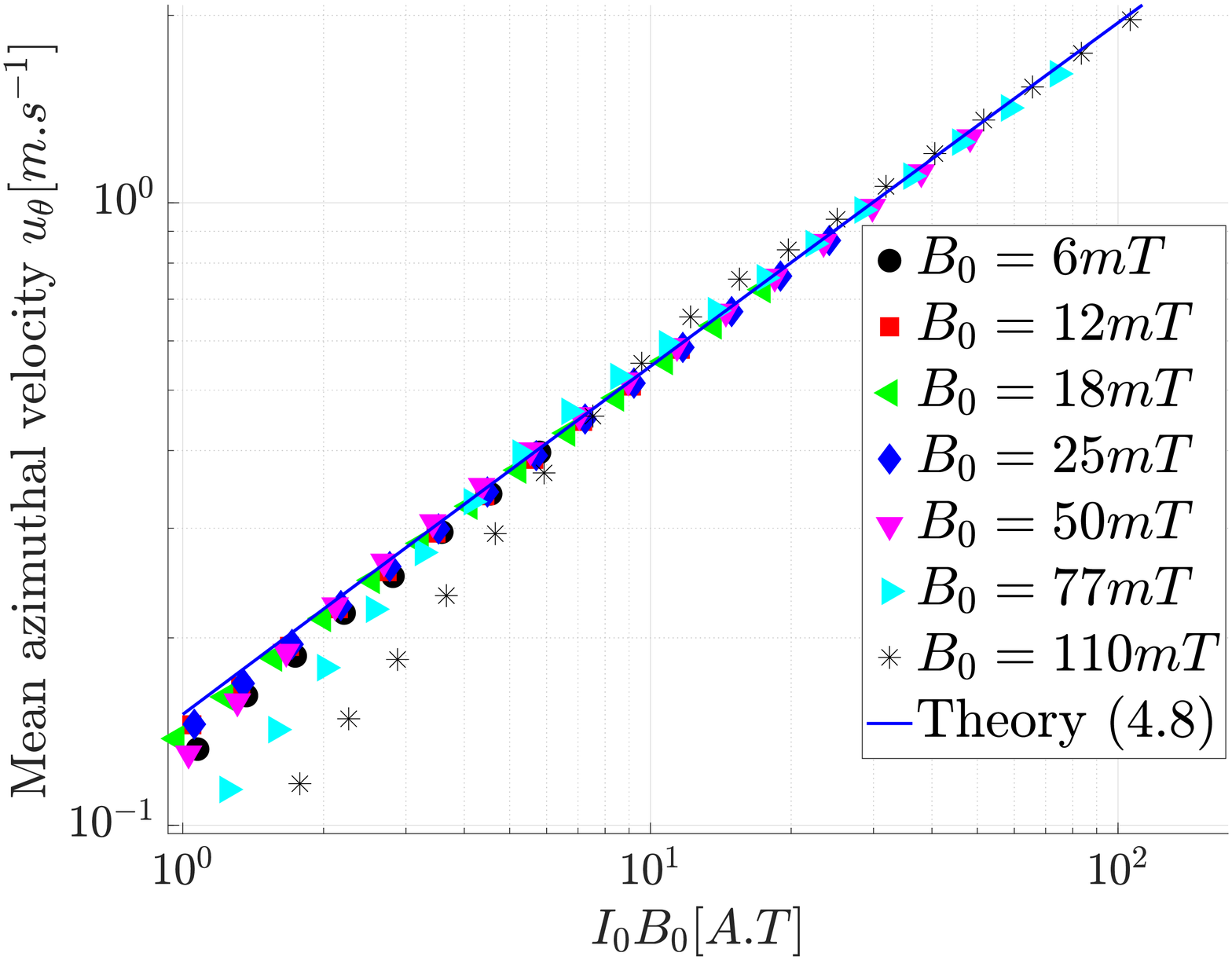}
    \includegraphics[scale=0.25]{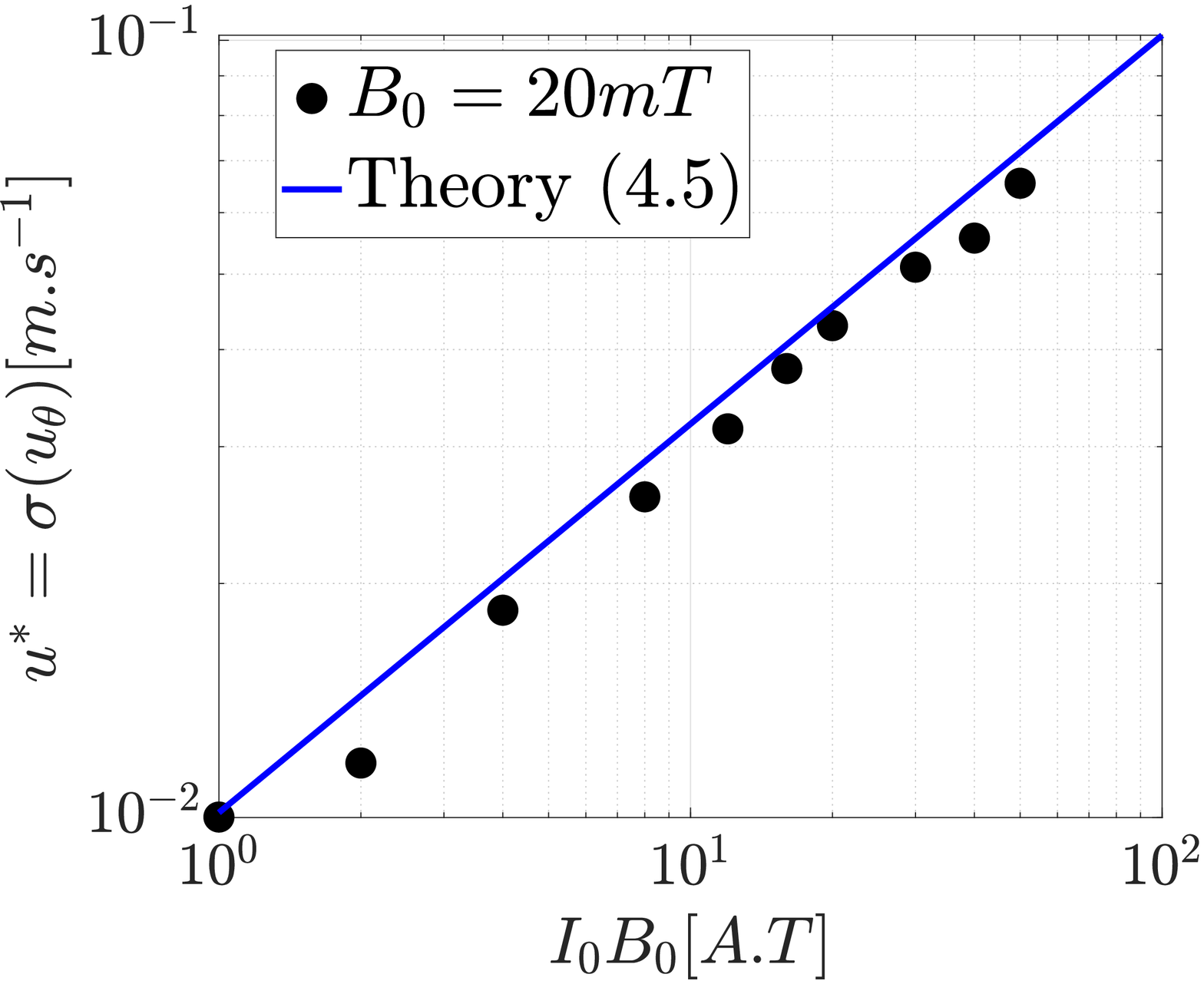}

    \caption{(a) Mean azimuthal flow $u_{\theta}$ vs the forcing $I_0B_0$ in the fully turbulent regime. All our data collapse on the theoretical prediction given by equation (\ref{ukepler}) (solid line). (b) Turbulent velocity $u^*$ estimated from the standard deviation of $u_\theta$ for $B=20$mT, and compared to equation (\ref{ustar}). $Ha\sim 0.57B_0$ with $B_0$ in $mT$.}
    \label{fig:scaling_kepler}
\end{figure}

Finally, when the forcing due to the Lorentz force becomes sufficiently strong, the flow undergoes a new transition from the previous laminar flows toward a presumably turbulent state. Fig.\ref{fig:scaling_kepler}(a) shows that for $I_0B_0>10$, all the data collapse on a single curve. The best power law to  fit the data is $u_\theta\propto (I_0B_0)^{0.54\pm0.01}$.

This result can be understood by assuming that the bulk flow is fully turbulent, such that the Reynolds stress gradient $\rho\partial_i\overline{ u_iu_j}$ overcomes the inertial terms involving the mean velocity $\rho (\overline{u_i}\partial_i){\overline u_j}$. In addition, because of the large  aspect-ratio of the experiment $h\ll r_m,\Delta r$, the Reynolds stress is mostly dominated by the components $\tau_{iz}=\rho\partial_z\overline{u_iu_z}$.
The azimuthal component of the Navier-Stokes equation therefore reduces to a  balance between the Reynolds stress $\rho\partial_z\overline{u_\theta u_z}$  and the applied Lorentz force:

\begin{equation}
u^*=\sqrt{\frac{I_0B_0}{4\pi \rho r}}
\label{ustar}
\end{equation}

\noindent where the additional factor $2$ comes from the estimate $\partial_z\overline{u_\theta u_z}\sim 2(u^*)^2/h$, with $u^*$ the typical velocity of the turbulent structures. This prediction for the intensity of the turbulent fluctuations matches very well our experimental data, as shown by Fig.\ref{fig:scaling_kepler}(b).

In the KEPLER experiment, the relevant boundary layers at the vertical wall are either the Hartmann layer at high $\Lambda$ or the B\"odewadt layer at low $\Lambda$. In the former, the stability of the layer is controlled by the Reynolds number $Re_H$ based on $\delta_H$, while in the latter, the relevant critical number is the Reynolds number $Re_B$ based on the thickness of the B\"odewadt layer.

\citet{Moresco04b} have predicted that pure Hartmann layers should become turbulent for $Re_H=Ha/Re>400$. Although it is well known that this criterion strongly depends on the Elsasser number when rotation effects are present~\citep{Davidson02, Moresco04b}, we expect a similar value here, since $\Lambda$ is never much larger than one in our experiment (except in the viscous-ideal regime where  $\Lambda\sim 10$).
In the regime studied here, as long as the forcing $I_0B_0$ remains larger than $10$, $Re_H$ is between $400$ and $2500$ and $R_B$ is larger than $200$. It is therefore reasonable to expect the Hartmann-B\"odewadt boundary layers to become turbulent at large forcing. Let assume a classical logarithmic profile for the turbulent boundary layer:

\begin{equation}
\frac{\overline{u_\theta}(z)}{u^*}=\frac{1}{\kappa}\ln(\frac{u^*z}{\nu}) +C
\label{log}
\end{equation}

\noindent where $\kappa$ is the Karman constant and $C\approx 5$. Fig.\ref{fig:scaling_kepler} shows that the turbulence intensity is relatively small, $u^*/\overline{u_\theta}\sim 3.10^{-2}\ll1$, so equation (\ref{log}) evaluated in the bulk ($z=h/2$) can be approximated to 

\begin{equation}
\frac{\overline{u_\theta}}{u^*}=\frac{\ln (Re_{h})}{\kappa}
\label{loglaw}
\end{equation}

\noindent where the Reynolds number is supposed large enough to overcome contributions from neglected terms. 
By combining relations (\ref{ustar}) and (\ref{loglaw}), we finally obtain the following prediction for the turbulent regime:

\begin{equation}
\overline{u_\theta}=\alpha\frac{\ln Re_h}{\kappa}\sqrt{\frac{I_0B_0}{4\pi\rho r}}
\label{ukepler}
\end{equation}

Note that because of the assumptions made during the derivation of this law, agreement between this prediction and experiments is only expected up to some pre-factor $\alpha$ of order one. In particular, MHD effects may very well modify the properties of the turbulent layer through the value of the parameters $\kappa$ and $C$. 

Despite these approximations, solution (\ref{ukepler}) with $\alpha=1.5$, represented by the solid blue line in Fig.\ref{fig:scaling_kepler}, matches remarkably well the experimental data. In particular, Fig.\ref{fig:scaling_kepler} shows that the logarithmic correction  is necessary  to fit the data, as it clearly increases the exponent of the power law.

Several remarks can be made on this prediction. First,  except for the logarithmic correction, it is independent of $h$. This is reminiscent of the turbulent nature of the boundary layers. This expression is only valid in the limit of a thin disc geometry, which leads to take $h$ rather than $r$ in the non-linear term. When $h$ is increased (experiments not reported here), we observed that this Keplerian profile (the radial scaling in $r^{-1/2}$) disappears. However, the square-root dependency in $I_0B_0$ stands because it comes from a natural balance between the advection and the Lorentz force \citep{Boisson17}. 
Note also that because this expression requires a destabilisation of the  boundary layers, it relies on a $3D$ cascade of energy. This forward transfer  of energy towards small  scales is however only required for scales smaller than the typical thickness $h$, and no constrains exists on the larger scales. This explains why the scaling law (\ref{ukepler})  can be  associated with  quasi-bidimensional turbulence. On the other hand, this regime is only observed for interaction parameter  not too large, namely $N_t<10$. It means that although quasi-bidimensional, it can not  be described by purely $2D$ Navier-Stokes equations.

Because the angular velocity profile is $\Omega\propto r^{-3/2}$,we refer to this solution as {\it Keplerian turbulence}. Although generated through a different mechanism, such profiles are expected in many astrophysical  objects in which  the centrifugal force balances gravity forces, as discussed in more details in section $5$.




\subsection{ Transitions between the different regimes}

\begin{figure}
    \centering
    \includegraphics[scale=0.4]{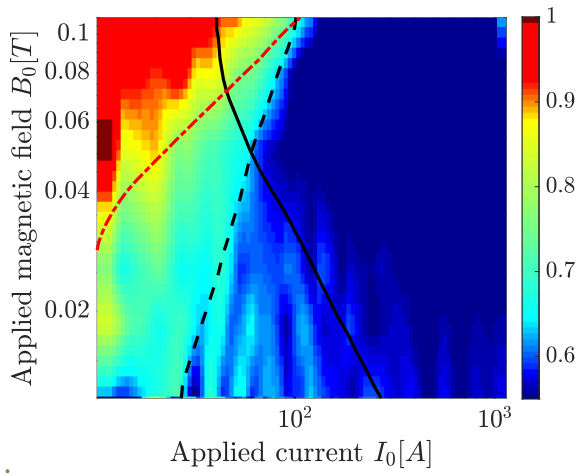}
    \caption{Colormap of the exponent $\xi$ of the mean azimuthal flow $u_{\theta}\propto I_0^\xi$ in the parameter space ($I_0,B_0$). The three regimes $\xi\sim 1$, $\xi\sim 2/3$ and $\xi\sim 1/2$ can be clearly distinguished (see the text).}
    \label{fig:exponents}
\end{figure}

Fig.\ref{fig:exponents} is very insightful to understand what controls the transition between the different regimes discussed above. To obtain this figure, we calculate the characteristic exponent of the scaling law 

\begin{equation}
\xi=\frac{\partial \ln (\overline{u_\theta})}{\partial \ln I_0}
\end{equation}

\noindent and plot it in the parameter space $(I_0,B_0)$. The red area corresponds to the {\it viscous-ideal} regime ($\xi\sim 1$), the light blue to  the {\it inertial-resistive} case ($\xi\sim 2/3$) and the dark blue to the  {\it Keplerian turbulence} ($\xi\sim 1/2$). The red dash-dotted line corresponds to $N_t=10$ and delimitates the region between strong and moderate magnetic field. Indeed, for large interaction parameter, most of the current passes through the thin Hartmann layers where the corresponding Lorentz force is balanced by viscosity (rather than inertia). As $N_t$ is reduced below $10$, the B\"odewadt boundary layer generates a strong radial inflow near the wall. The interaction between this wind and the inertia in the bulk leads to the $2/3$ scaling law, as explained in the previous section. Note that in this regime, the Reynolds number $Re$ is sufficiently small so the flow remains laminar. Alternatively, this transition can be predicted by balancing the two expressions for the mean azimuthal velocity given by \ref{viscous_ideal} and \ref{inertial_resistive}, leading to the (velocity-free) dimensionless ratio $4\pi r^2B_0^2\sqrt{\nu\sigma^3}/I_0\sqrt{\rho}$. The transition between the \textit{viscous-ideal} and the \textit{inertial-resistive} regime then occurs when this ratio is equal to one. Note that this ratio is an Elsasser number $\Lambda$ in which the velocity is evaluated thanks to equation \ref{viscous_ideal}.

\begin{figure}
    \centering
    \begin{subfigure}{0.49\textwidth}
        \includegraphics[width=\textwidth]{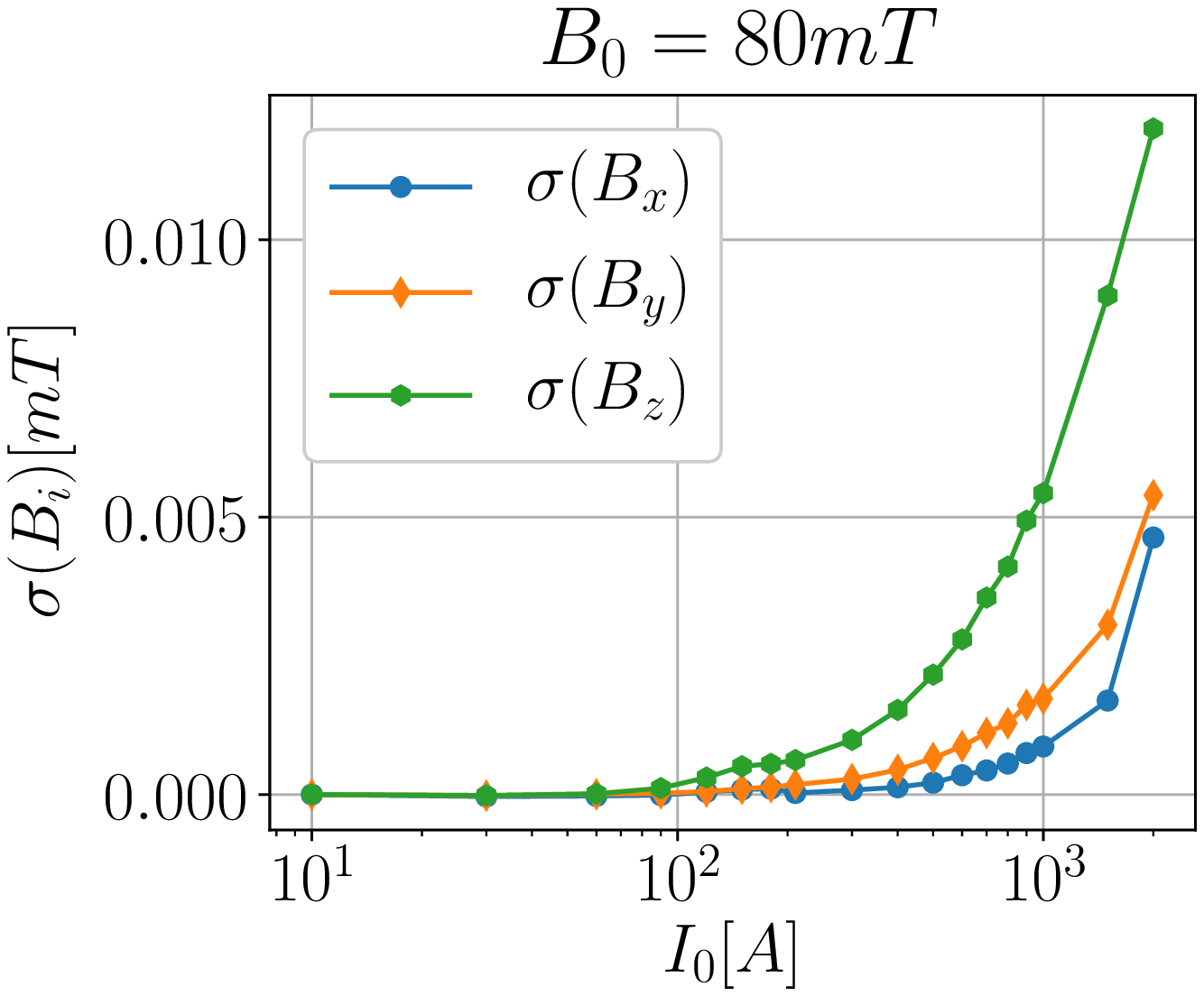}
        \caption{}
        \label{fig:std_B}
    \end{subfigure}
    \hfill
    \begin{subfigure}{0.49\textwidth}
        \includegraphics[width=\textwidth]{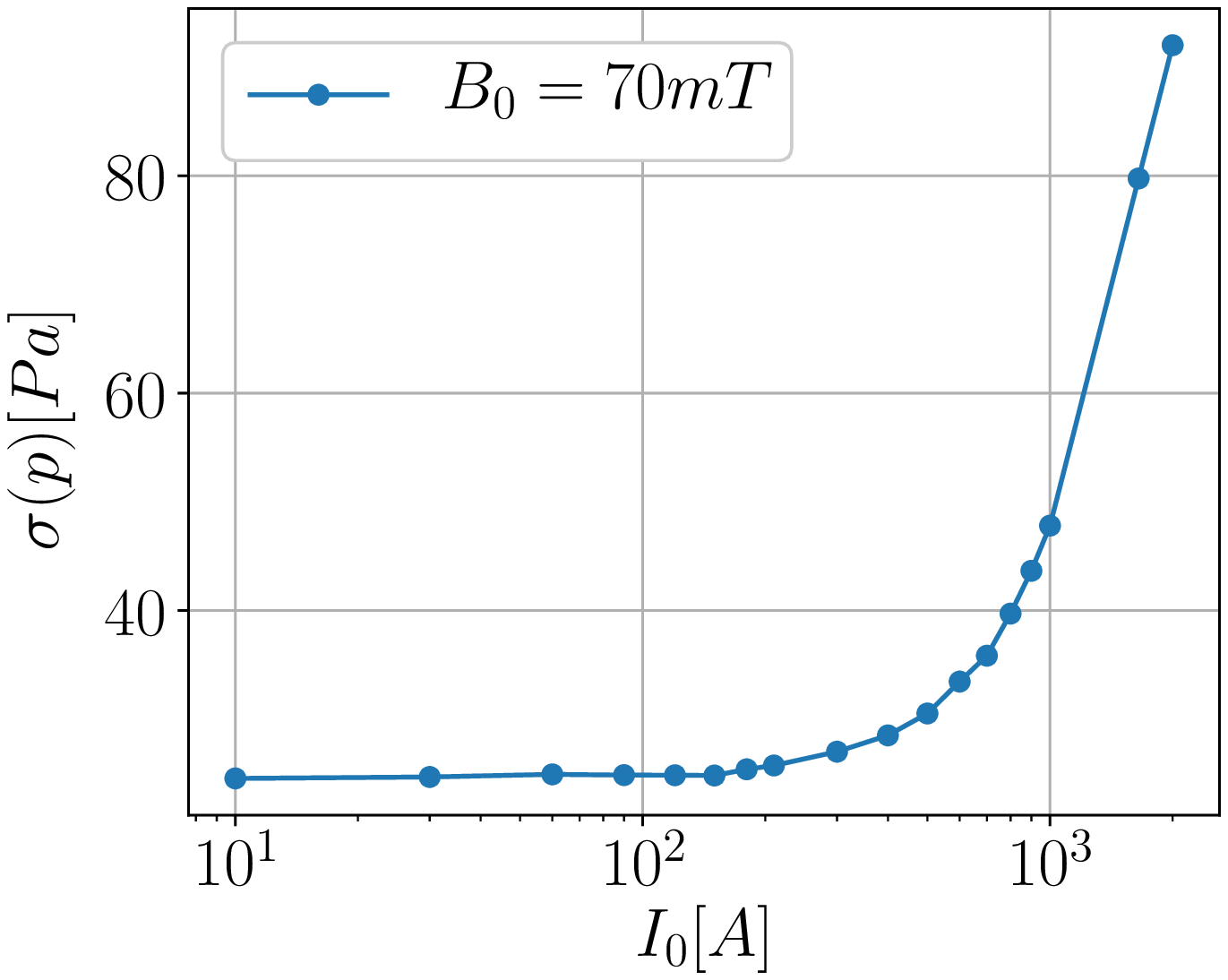}
        \caption{}
        \label{fig:std_p}
    \end{subfigure}
    \caption{(a) Fluctuations (standard deviation) of the three components of the magnetic field as a function of the injected current $I$ for an imposed magnetic field $B_0=800G$ ($80mT$). (b) Fluctuations (standard deviation) of the pressure as a function of the injected current $I$ for an imposed magnetic field $B_0=700G$ ($70mT$).}
\end{figure}

The transition from the $2/3$ regime to the turbulent one is less trivial. At small magnetic field, the B\"odewadt layer is much smaller than the Hartmann one, and its stability is fully controlled by a critical  Reynolds number $Re_B=u\delta_B/\nu$ based on the B\"odewadt layer thickness (black solid line). 
In contrast, at large magnetic field,  the stability of the thinner Hartmann layer is measured by the Reynolds number $Re_H$ based on the thickness of this layer.  We observe that the flow consequently bifurcates to the Keplerian turbulent regime along the line $Re_B\sim 400$ (solid black line) at low field, while the transition rather takes place at a critical $Re_H\sim 300$ (dashed black line) at large field (for $B_0>50mT$).  Fig.\ref{fig:exponents}  therefore strongly suggests that the transition to the Keplerian turbulence is therefore ultimately triggered by the instability of Hartmann-B\"odewadt layers.

Figures \ref{fig:std_B} and \ref{fig:std_p} show that the transition to turbulence in  the experiment is associated with a strong  (yet  continuous) increase of the fluctuations of both induced magnetic field and pressure, without hysteresis. These  results suggest that  the  Keplerian turbulence always occurs through a smooth supercritical transition.

\subsection{Energy budget}

Because conducting fluids can either dissipate energy viscously or by ohmic dissipation, the question of the energy budget of such electromagnetically-driven flows is highly non-trivial. In electromagnetically-driven flows, the energy equation is \citep{Reddy18}:
\begin{align}
    \partial_tE_u = -D_{\nu} + T \\
    \partial_tE_b = -D_{\eta} - T + F
    \label{eq:bilan}
\end{align}
where $E_u = \rho\int_Vd^3x\frac{\bm{u}^2}{2}$, (respectively $E_b = \int_Vd^3x\frac{\bm{B}^2}{2\mu_0}$) is the total kinetic (resp. magnetic) energy, $D_\nu=\rho\nu\int_Vd^3x(\bm{\nabla}\times\bm{u})^2$ (resp. $D_\eta=\frac{1}{\sigma}\int_Vd^3x\bm{j}^2$) is the viscous (resp. Ohmic) dissipation. $T=\frac{1}{\mu_0}\int_Vd^3x\bm{B}\otimes\bm{B}:\bm{\nabla}\otimes\bm{u}$ is the term which couples the two equation and corresponds to the transfer of energy between the two fields and $F=\eta\oint_{\partial V}\bm{dS}\cdot(\bm{j}\times\bm{B})$ is the injected power.

In addition, one can also define the efficiency $\gamma$ of the transformation as the ratio between the work of the Lorentz force and the total injected power $\gamma = T/F$. In stationary state, this measure of the amount of energy which is transfer to the kinetic part can also be written $\gamma = D_\nu/(D_\nu + D_\eta)$. This energy budget essentially depends on the flow regime, or more precisely on the degree of turbulence in the flow. First, in the \textit{viscous-ideal} regime, the velocity gradients are essentially confined to the two top and bottom Hartmann boundary layers and the electrically current almost entirely passes through these layers. At the leading order, the current density is therefore given by $j\sim \frac{I_0}{4\pi r \delta_{Ha}}$, and the dissipations are given by
\begin{align}
    D_\nu \sim \frac{2\rho\nu S U^2}{\delta_{Ha}} \\
    D_\eta \sim \frac{2SI_0^2}{(4\pi r )^2 \delta_{Ha}}
\end{align}
where $S$ is the surface of the endcap. In the laminar regime, the velocity $U$ is given by expression (\ref{viscous_ideal}), and the dissipation ratio $D_\nu/D_\eta$ is simply equal to one. The efficiency $\gamma$ of the magnetic to kinetic energy conversion is therefore $50\%$.

An exact calculation is not possible in the turbulent regime, but a general trend can be obtained: our scaling laws in this regime suggest that most of the viscous dissipation occurs in the Bodewadt boundary layer of typical size $\delta_\Omega$, and that the electrical current flows through the entire height of the channel. At leading order one would get $j \sim \frac{I_0}{2\pi r h} $ and:
\begin{align}
    D_\nu \sim \frac{\rho\nu S U^2}{\delta_\Omega}\sim \frac{\rho\nu S U^{5/2}}{\sqrt{\nu r}}\\
    D_\eta \sim \frac{SI_0^2}{\sigma 4 \pi^2 r^2 h}
\end{align}
This leads to the dissipation ratio: $\frac{D_\nu}{D_\eta} \sim 4\pi^2r^{3/2}\nu^{1/2}h\sigma \frac{U^{5/2}}{I_0^2}$ where the prefactor is equal to $0.1$ in our setup. By setting $U\sim 1m.s^{-1}$ and $I_0\sim 1000A$ as typical values of the turbulent regime, one gets $D_\nu/D_\eta \sim 10^{-7}$. In other words, the turbulence "destroy" the efficiency of the energy transformation, and the dissipation in the turbulent regime is almost entirely due to Ohmic dissipation, thus leading to a strong Joule heating of the liquid metal at large current \textit{and} large magnetic field. As explained in section 2, the inner cylinder is therefore cooled in order to keep the experiment at variations of $\sigma$ less than one percent.

These results of an efficiency bounded by $50\%$ and decreasing with the level of turbulence are surprisingly similar to the one obtained by \citep{Reddy18} in a very different MHD setup, suggesting some universal behavior of the energy budget of electromagnetically-driven flows.

\subsection{Low frequency oscillations}

At large magnetic field and moderate current, there is a small region of the parameter space in which both $Re_B$ and $N_t$  are beyond their critical values $Re_B^c$ and $N_t^c$. Interestingly, in this region, the velocity  field exhibits slow dynamics characterized by low frequency oscillations and interpreted as large scale structures being  advected by the mean flow.

\begin{figure}
\centering
           \includegraphics[width=0.9\textwidth]{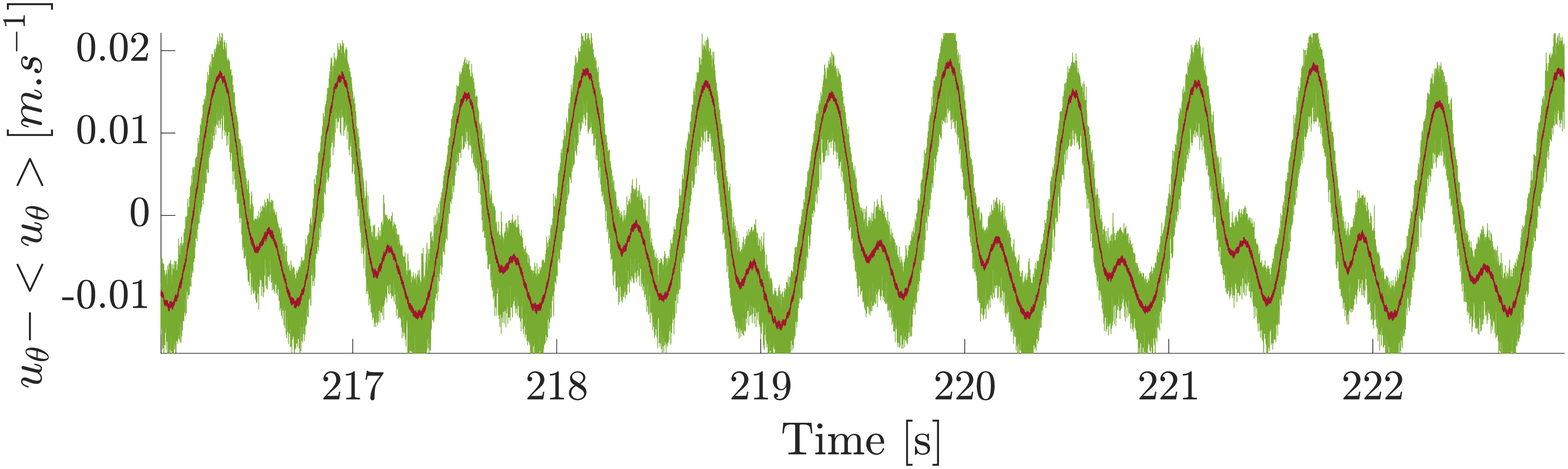}
            \includegraphics[width=0.4\textwidth]{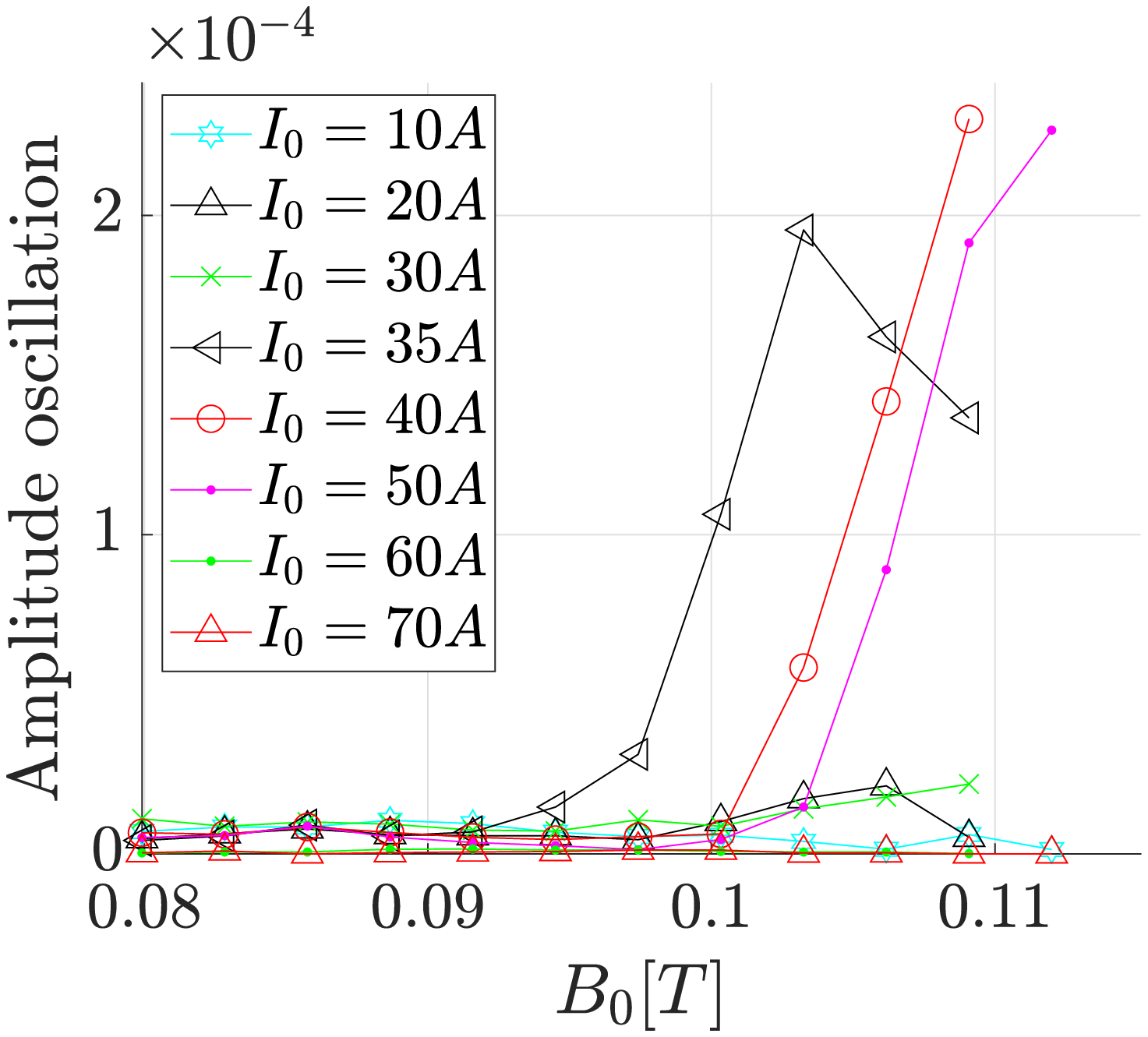}
             \includegraphics[width=0.4\textwidth]{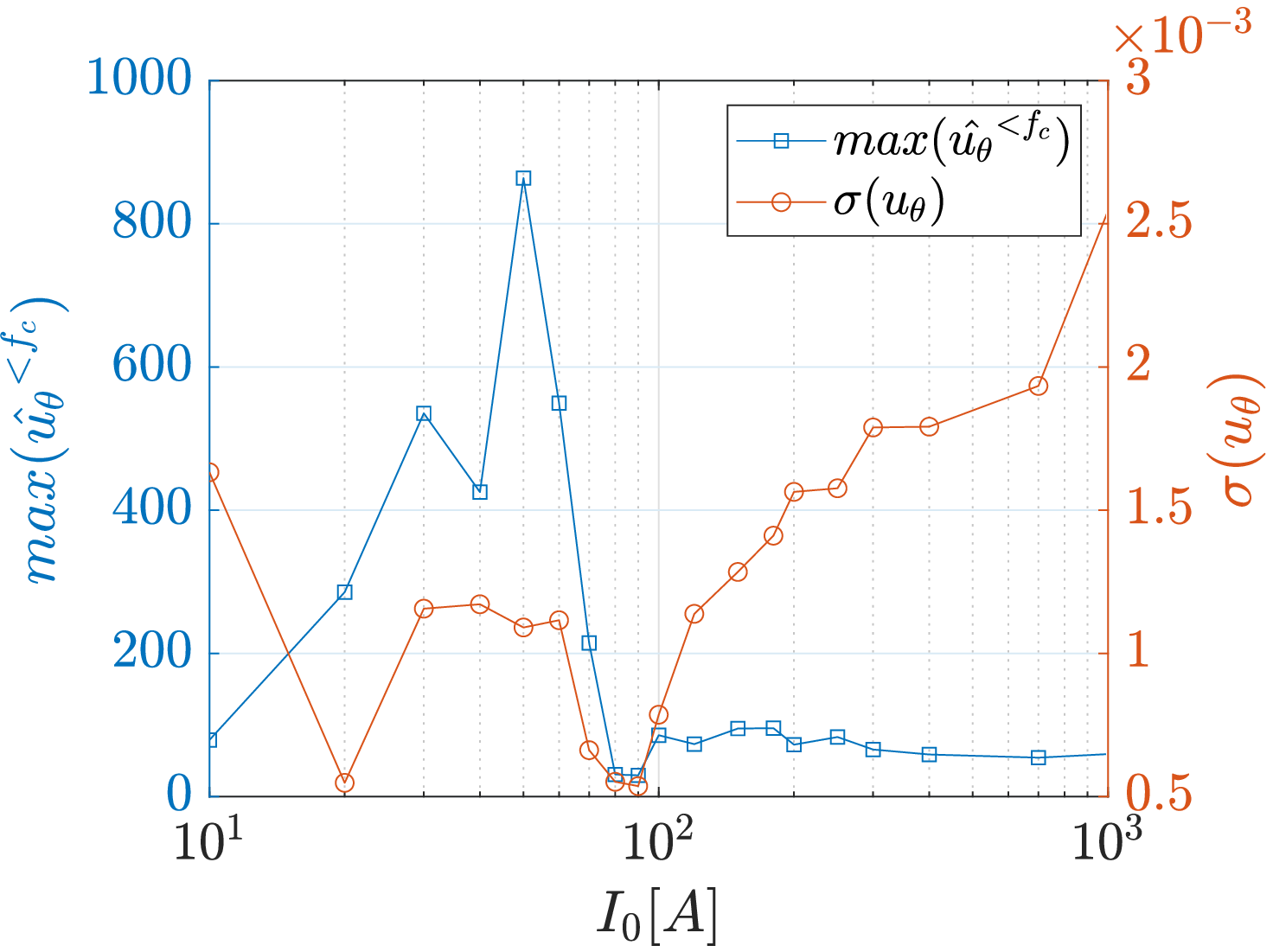}

    \caption{(a) Time series of the fluctuations of $u_\theta$, measured with a potential probe for $B_0=110mT$ and $I=30A$. (b) Amplitude of the low frequency oscillation as a function of the magnetic field $B_0$ for different values of $I_0$. (c). Same as a function of  the  current, together with the standard deviation of the azimuthal component of the velocity.}
    \label{time_osc}
\end{figure}

Indeed, for  $B_0>95mT , 30A<I_0<100A$, the frequency spectra generally show a peak of energy at low frequencies ($f\sim 1-5 Hz$).  Time series of Fig.\ref{time_osc}(a)  show that this condensate of energy at large scale corresponds to a periodic oscillation of the flow with its lowest frequency at $f\sim 2Hz$, quite comparable to the advection time $t_u=\Delta r/u_{\theta}\sim 5.10^{-1}$s. As shown by Fig.\ref{time_osc}(b), these oscillations appear only at a critical current $I_0\sim 30A$, corresponding to $Re_B>Re_B^c=400$. The transition to  this regime takes the form of a supercritical bifurcation which occurs only in a given range of current with no sign of hysteresis.  As the current is increased at fixed magnetic field, the amplitude of these oscillations suddenly goes to zero (for $N_t\sim10$) and are replaced  by the  turbulent flow described above (see Fig.\ref{time_osc}(c)).
   
These results are very close to the ones obtained by \citet{Messadek02}, who reported similar low frequency oscillations. In their experiment, the current was injected by an array of electrodes located at an intermediate radius between the inner electrode and the outer cylinder of the experiment. These oscillations were therefore attributed to an instability of the shear layer generated near the electrodes between the active flow and the fluid at rest in the outer part of the experiment. There is however no steady region in our experiment, since the current is injected directly between the inner and the outer cylinder. These oscillations are more likely to be the consequence of some boundary layer instability. In this perspective, it is interesting to note that the Reynolds number based on the thickness of Shercliff layers generated at the radial boundaries is also relatively large ($Re_{sh}>10^3$), such that an instability of these layers could be responsible for the oscillations. In this regard, these results are much closer to  the periodic oscillations found by Tabeling near the onset of turbulence in electromagnetically-driven Taylor Couette flows \citep{Tabeling81}.

In any case, these oscillations can be attributed  to large scale bi-dimensional structures of typical size $l\sim 6 cm$ which are advected  by the mean azimuthal flow. Because they occurs in the region  delimited by $(N_t>10, Re>9000)$, they could in principle be described  by purely $2D$ Navier-Stokes equations.

\section{Velocity  profiles and transport of angular momentum}

\begin{figure}
    \centering
    \includegraphics[scale=0.20]{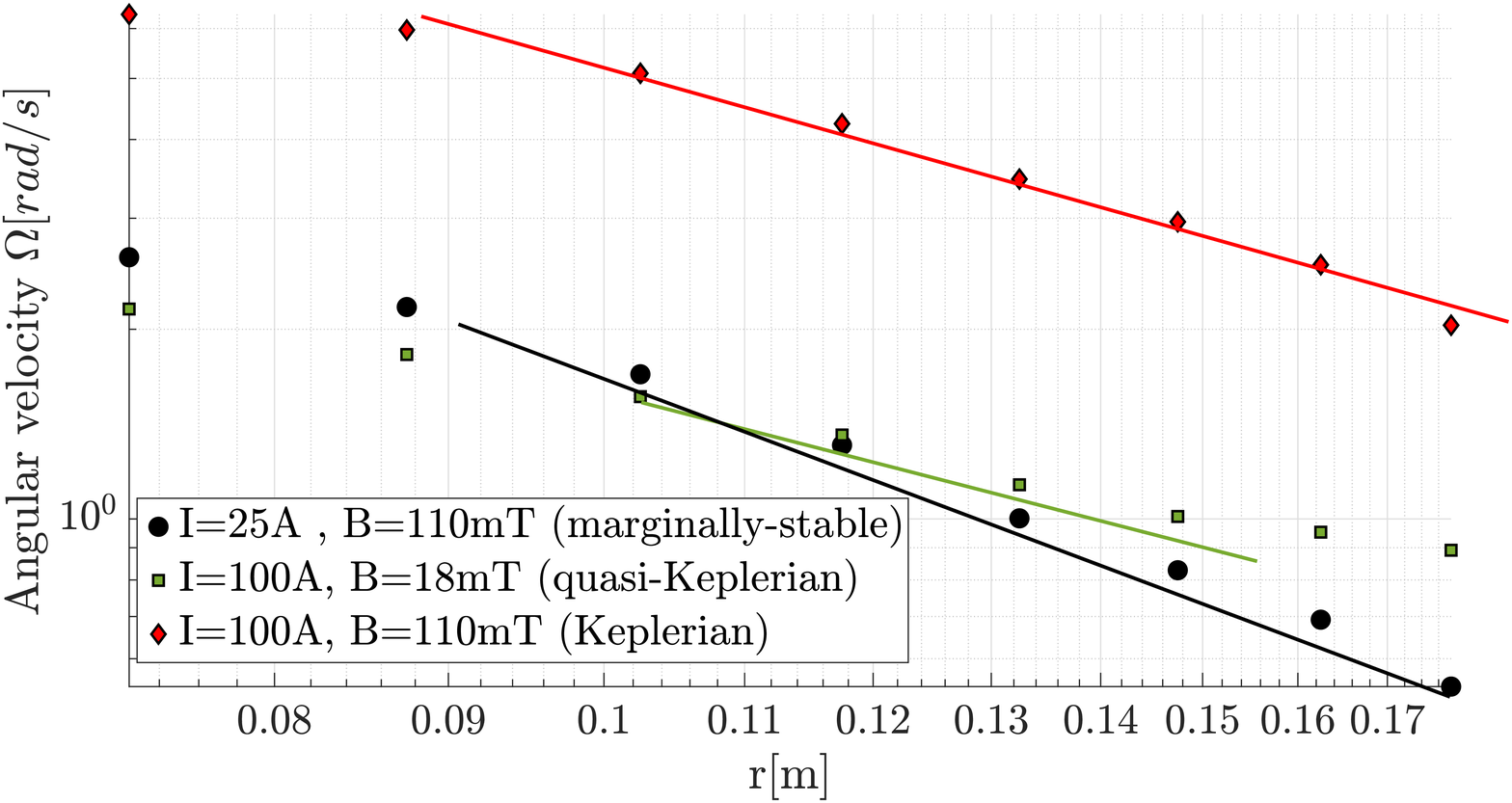}
    \includegraphics[scale=0.25]{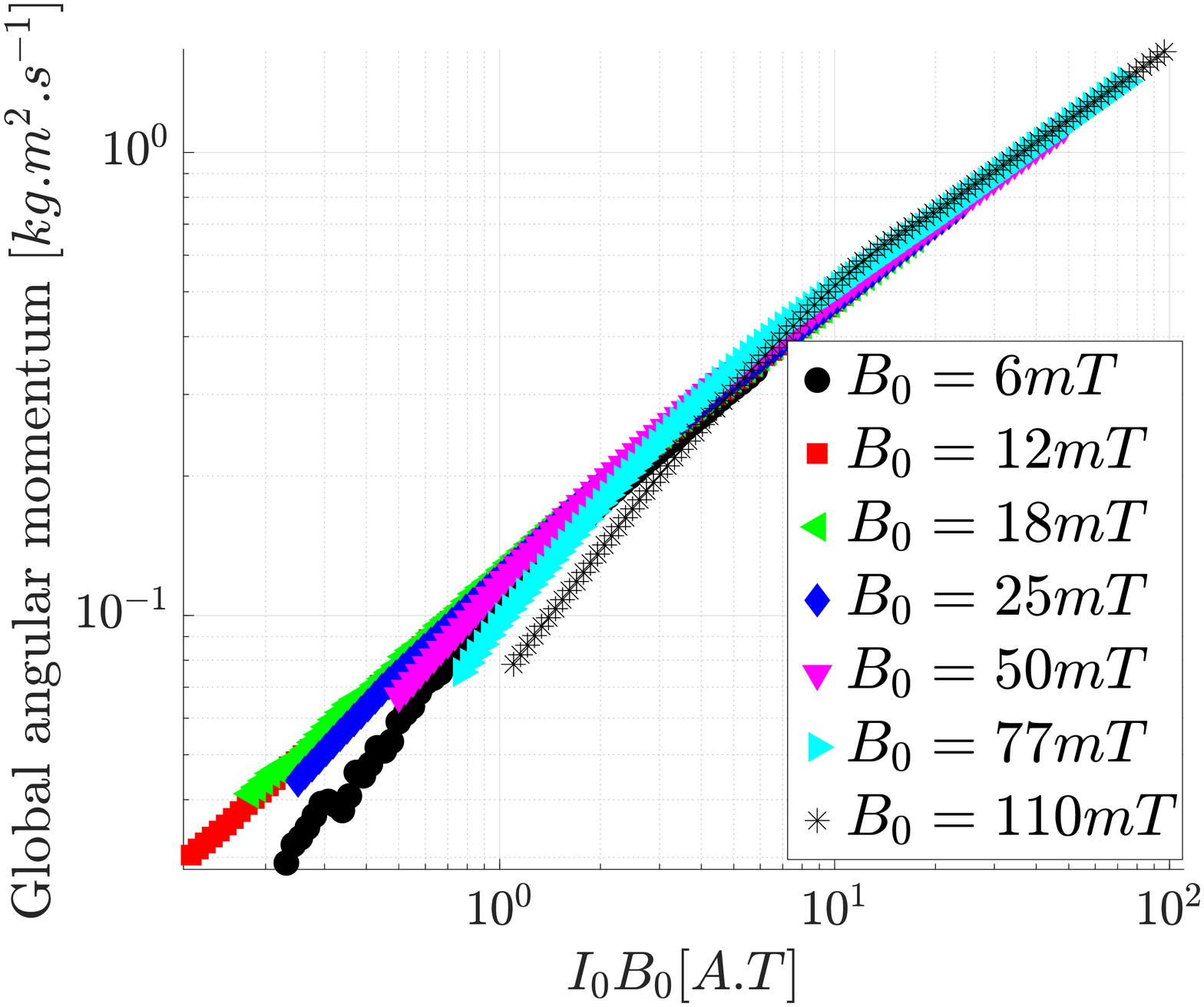}    
        \caption{(a) Angular velocity $\Omega(r)$ obtained from potential probes, for three typical set of parameters ($I_0,B_0$), corresponding to the three different regimes predicted. (b) Global angular momentum versus $I_0B_0$. $Ha\sim 0.57B_0$ with $B_0$ in $mT$.}
    \label{3profiles}
\end{figure}

Expressions (\ref{viscous_ideal}), (\ref{inertial_resistive}) and  (\ref{ukepler}) lead to three different predictions for the velocity profile. Interestingly,  each of these profiles describes a well-known  physical situation encountered when studying the stability of toroidal flows. Figure \ref{3profiles}(a) shows the angular velocity $\Omega$ for three different sets of parameters corresponding to these three regimes. 

First, the black circles correspond to the {\it viscous-ideal} profile $\Omega\propto 1/r^2$ which generates an angular momentum $l=\Omega r^2$ constant in radius. This profile is therefore marginally stable, since $l=$cste is precisely the  Rayleigh's criterion for the stability of inviscid circular flows. Except close to the inner cylinder, the data are relatively close to the prediction. Note that the transport of angular momentum is then achieved by the viscous dissipation in the boundary layer.

The {\it inertial-resistive} regime (green squares) exhibits a less steep profile, as predicted by equation (\ref{inertial_resistive}). This type of profile  between  global  rotation and marginal stability (in which $q=-\frac{\partial \ln\Omega}{\partial \ln r}$ is between $0$ and $2$) is generally called a quasi-Keplerian profile. However, if the theory succeeds in predicting the magnitude of the flow, the predicted profile $\Omega\propto r^{-4/3}$  is only observed for a small range of $r$ in the middle of the gap. This is not  surprising,  as  the unknown function $f(\Delta r/r)$ in the prediction (\ref{inertial_resistive}) must strongly modify the velocity profile.

Finally, the {\it turbulent regime} exhibits a large range on which the velocity field satisfies the predicted Keplerian rotation rate, with $q=3/2$.  Such keplerian profiles are very different from what is observed in Taylor-Couette flows at large Reynolds number. In the latter, once both the bulk flow and the radial  boundary layers are turbulent, the differential rotation remains generally confined close to the radial boundaries, and the bulk flow exhibits a flat profile. The persistence of  a mean Keplerian profile $u_\theta\propto \frac{1}{\sqrt{r}}$ in the bulk flow comes from the volumic Lorentz force driving  the flow, absent in boundary-driven TC flows\citep{Avila12,Avila17}.

\begin{figure}
    \centering
    \includegraphics[scale=0.40]{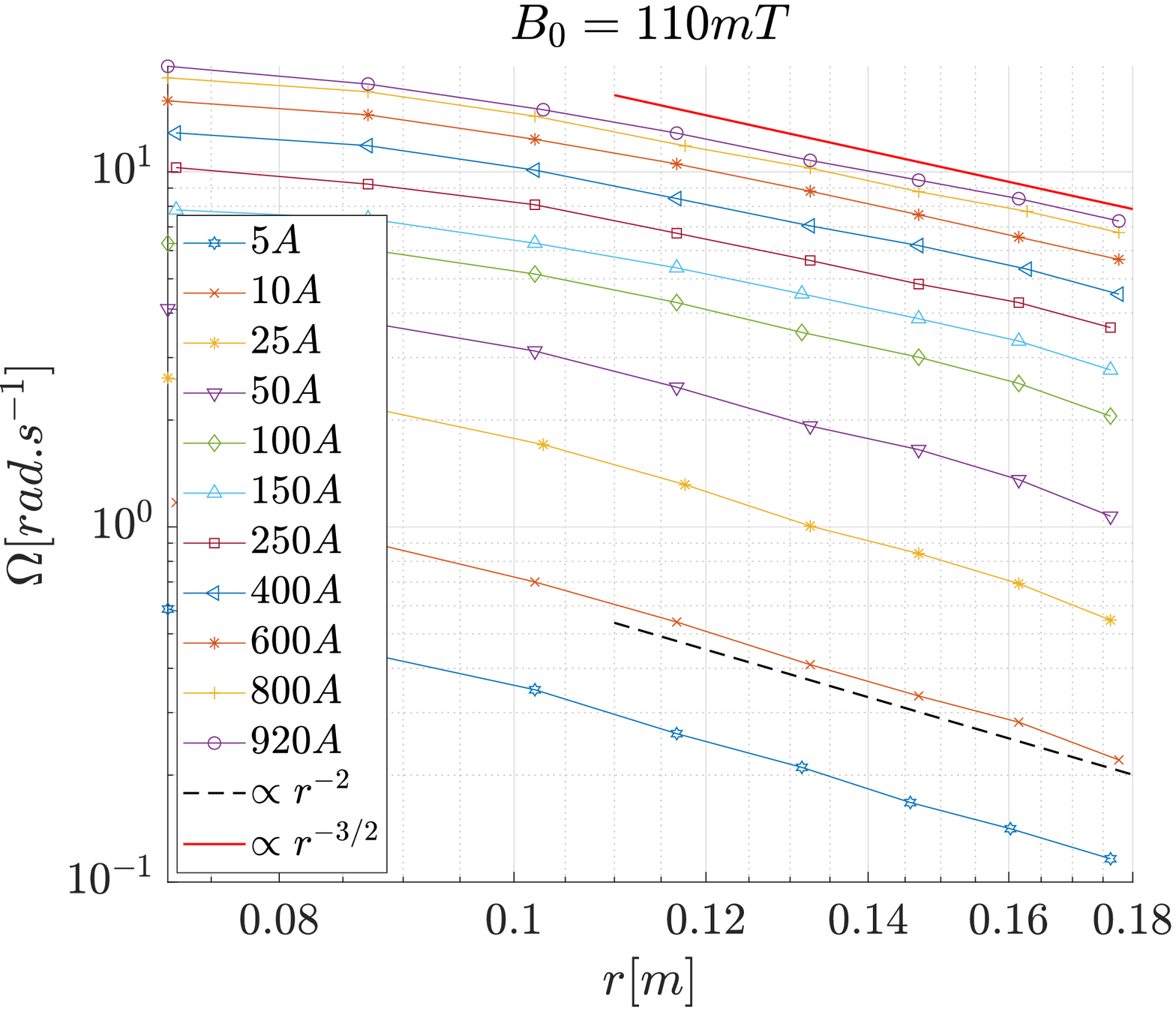}
        \includegraphics[scale=0.45]{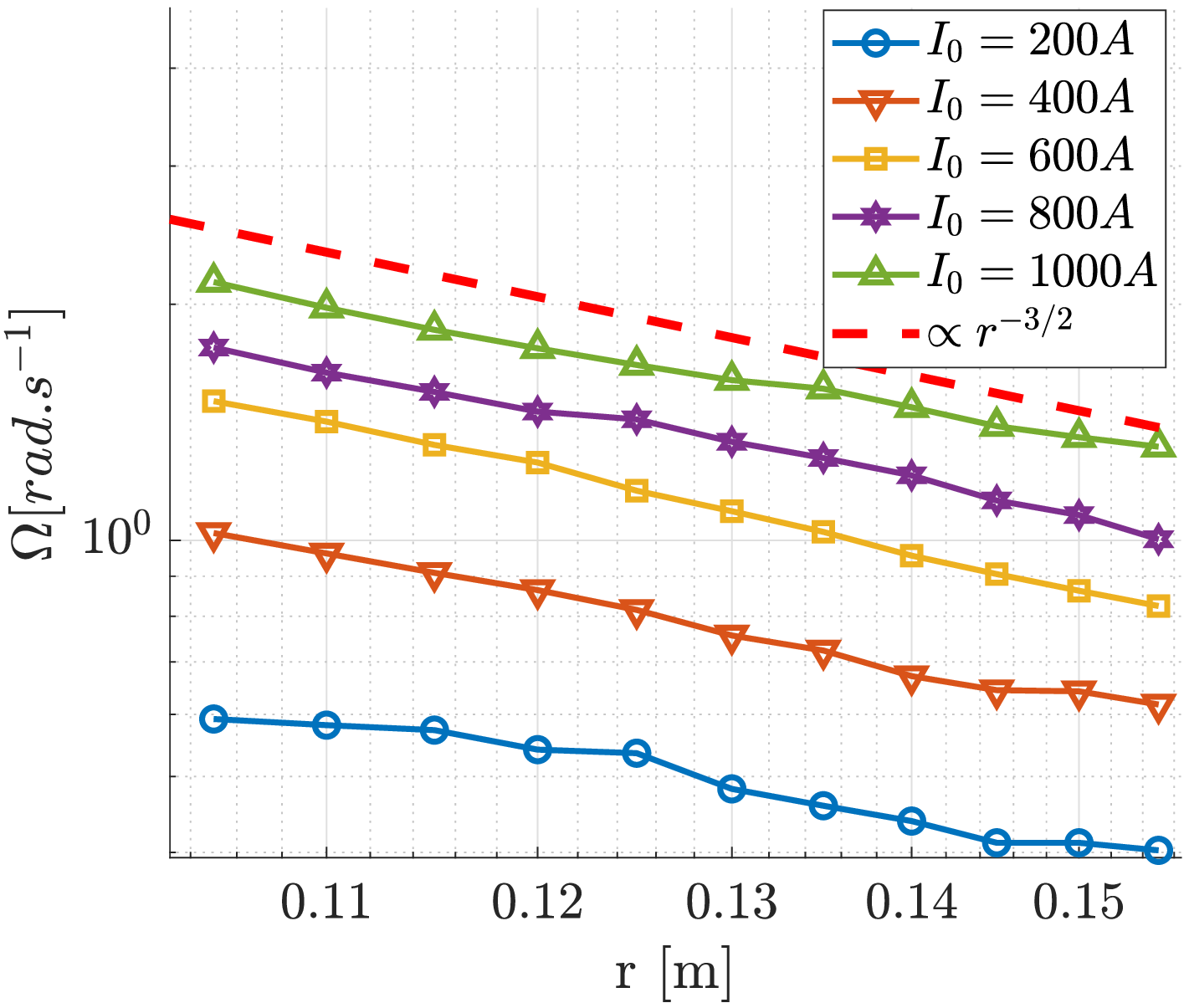}
    \caption{(a) Angular velocity obtained from potential probes for $B_0=110mT$, for increasing current showing the transition from the laminar profile (\ref{viscous_ideal}) (dashed black line) to Keplerian turbulent flows (\ref{ukepler}) (solid red line). (b)
Angular velocity $\Omega(r)$  obtained from Doppler probes, for five values of injected current $I$ ($200A$,$400A$,$600A$, $800A$ and $1000A$). The dashed line corresponds to a power law $r^{-3/2}$. Here $B_0=80mT$. $Ha\sim 0.57B_0$ with $B_0$ in $mT$.} 
    \label{fig:profils}
\end{figure}

Fig.\ref{fig:profils}(a) (resp. Fig.\ref{fig:profils}(b) ) shows radial profiles of the  mean azimuthal velocity field obtained from potential probes (resp. Doppler velocimetry). First, the good agreement between the two types of measurement indicate that at least in this regime, the vertical component of the velocity is negligible. In addition, it shows that the change in the velocity profile from marginal flows to Keplerian rotation rates appears smoothly when the current is increased at fixed magnetic field. Moreover, Fig.\ref{fig:profils}(b) focuses on the Keplerian regime only, and shows that once in the turbulent regime, the velocity profile remains completely unchanged (the current has been varied by an order of magnitude without noticing any change). This suggests some {\it ultimate} nature of this turbulent regime.\\

The term {\it ultimate regime} used here refers to the ultimate regime of thermal convection predicted by \citet{Kraichnan62}, who used similar arguments to derive a law for heat transport in strongly turbulent flows. 
Because of the  analogy between Rayleigh-Benard convection and Taylor-Couette flows, an analogue  ultimate regime has been extensively studied in rotating flows in cylindrical geometry, see for instance \citep{Huisman12}. 
The evolution of the global angular momentum 

\begin{equation}
L=\int_{r_{min}}^{r_{max}} r^2\overline{u_\theta}(r) dr
\end{equation}

\noindent is shown in Fig.\ref{3profiles}(b) and clearly illustrates the existence of three different types of angular momentum transport. As in Taylor-Couette flows, the Keplerian regime can be regarded as an ultimate turbulent regime occurring only at very large Reynolds number and transporting very effectively the angular momentum outward. 

This question of the angular momentum transport is crucial in astrophysics. As explained in the introduction, accretion discs around black holes and proto-stars are known to exhibit similar turbulent rotation profiles $\Omega\propto r^{-3/2}$ and exhibit very efficient angular momentum  transport for reasons that are still unclear.  Keplerian rotation rates can only be approached in Taylor-Couette experiments, and there is a fairly active debate as to whether or not such quasi-Keplerian flows are turbulent and efficient in transporting angular  momentum at large Reynolds number \citep{Goodman06,Eckhardt07,Paoletti11,Paoletti12,Avila12,Fromang19}.  It is therefore interesting to obtain here a truly Keplerian flow, in a regime undoubtedly  turbulent. Although the turbulence originates from mechanisms radically different from its astrophysical counterpart, the KEPLER experiment provides an interesting experimental analogue of a Keplerian accretion disc, at least regarding  the angular momentum profiles, the aspect ratio and the presence of a magnetic field which are typical of these astrophysical objects.  Naturally, the problem is quite different from the Taylor-Couette problem, mostly because of the presence of the volumetric Lorentz force driving the flow and the fact that the mean flow is Rayleigh-stable. In particular, the KEPLER experiment is associated with a transverse current of azimuthal motion which is not conserve in the radial direction, contrary to classical setups. In this regard, the transport of angular momentum in our system is therefore more analogue to Rayleigh-Benard convection with internal heating. A full discussion of this analogy and of the angular momentum transport in this experiment is thus beyond the scope of this paper, and will be reported in a forthcoming article.

\section{Quasi-bidimensional turbulence}

\subsection{Power spectrum and energy fluxes}

The above results show that at large forcing $I_0B_0$, the flow is clearly turbulent. The dimensionality of this flow is however less trivial. Most previous experiments on similar electrically-driven flows reported the generation of quasi-bidimensional turbulent flows at large magnetic field.  There are however two important differences with the results reported here: our experiment operates at a significantly smaller interaction parameter $N_t$, and takes place in a very thin layer of thickness $h\ll \Delta r$, leading to a different regime.

\begin{figure}
\centering
           \includegraphics[width=0.7\textwidth]{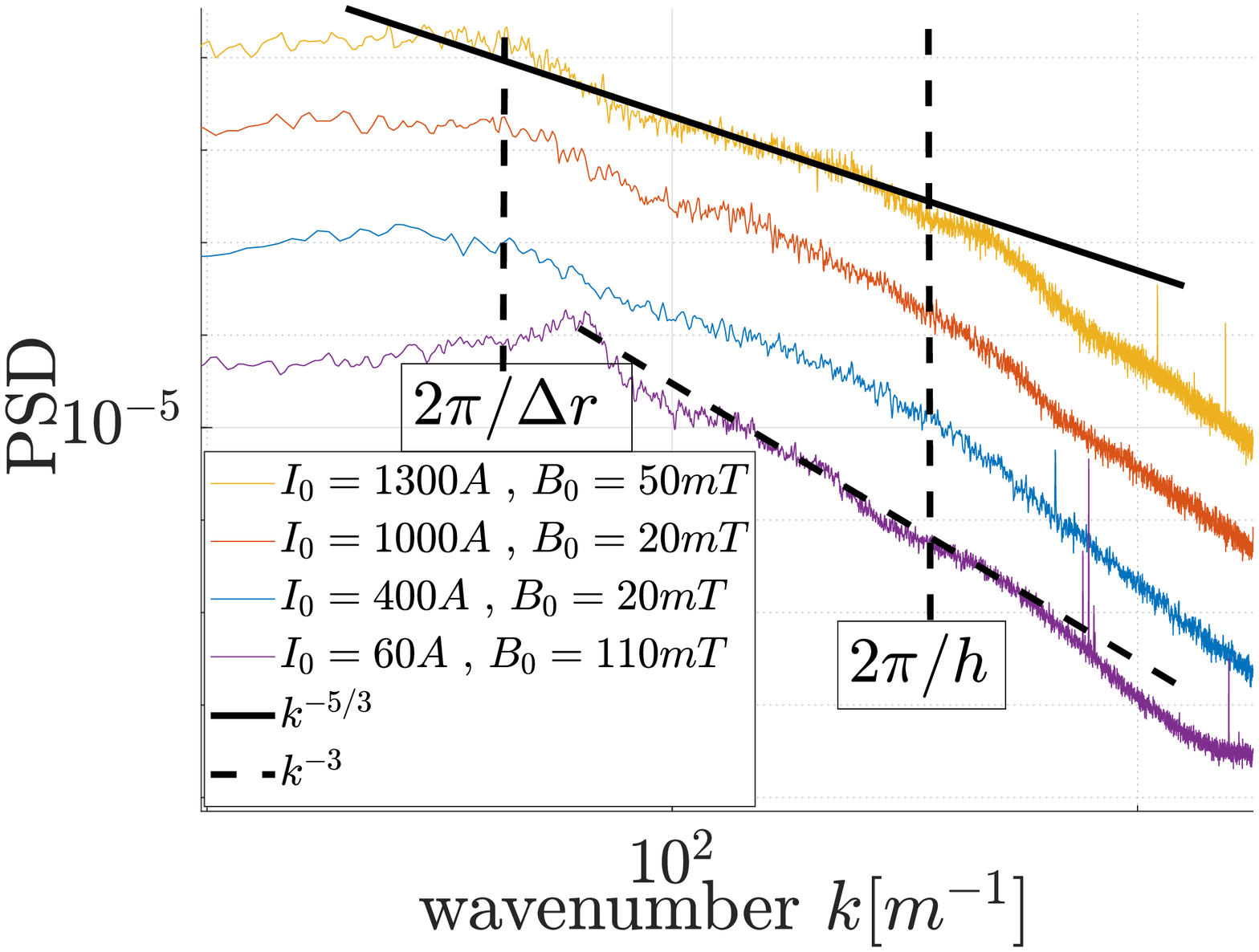}
               \includegraphics[scale=0.4]{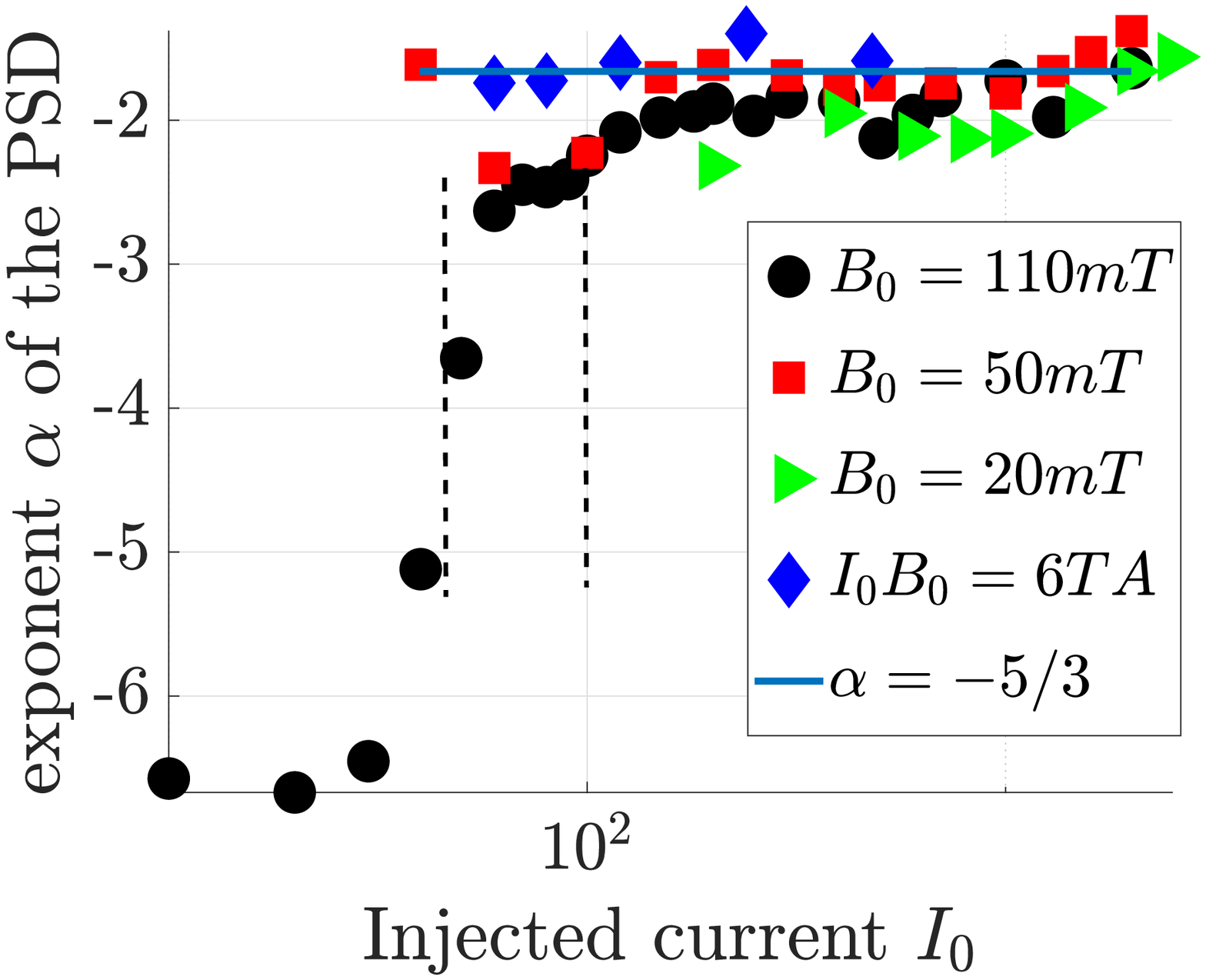}
           
    \caption{ (a) Power spectral densities measured with the potential probes at four different points in the parameter space. The lines correspond to scalings $E(k)\propto k^{-5/3}$ and $E(k)\propto k^{-3}$. (b) Coefficient of the slope of the power spectral density as a function of the injected current $I$ in the range: $k=[100-300]$. $Ha\sim 0.57B_0$ with $B_0$ in $mT$.}
    \label{spectrum}
\end{figure}

Fig.\ref{spectrum}(a) shows  typical examples of power spectra obtained in the turbulent regime. These spatial spectra were obtained from time series using a Taylor hypothesis based on the fact that the mean azimuthal flow  is much larger than the fluctuations. Note that the four cases studied here are indicated by circles in Fig.\ref{fig:map_patch}.

In three of the four cases presented in \ref{spectrum}, $Re>1,5.10^5$, $R_B>360$ and $\Lambda < 1$, such that they all belong to the turbulent regime  described in the previous section, but the interaction parameter $N_t$ remains between $0.25$ and $16$. Such values are much smaller than the ones used previously in similar experiments, but sufficiently large for quasi-2D turbulence to be generated.

In all cases except for the low frequency region, the velocity field $u_\theta$ exhibits important turbulent fluctuations ($u_\theta'/\overline{u_\theta}\sim 10\%$) and all energy spectra computed at $B_0<100 mT$ follow closely $E(f)\propto f^{-5/3}$ for $2\pi/\Delta r<k<2\pi/h$ and a much steeper slope for $k>2\pi /h$. The latter can probably be attributed to the cutoff of the potential probes, which are limited by the finite distance $l_p$ between the two electrodes measuring the local voltage and generate a $k^{-11/3}$ spectrum for wave numbers $k>2\pi/l_p$  \citep{Fauve98}.   Fig.~\ref{spectrum}(b)  reports the slope $\alpha$ of the PSD computed for wave numbers $k=[100-300]$, which correspond to distances midway between the largest (horizontal) scale $\Delta r$ and the smaller vertical scale $h$ of the experiment. It shows that this $k^{-5/3}$ exponent is observed for a large range of the parameters ($I_0,B_0$), and deviations from it occur only at very large field and small current.

\begin{figure}
    \centering
    \includegraphics[scale=0.35]{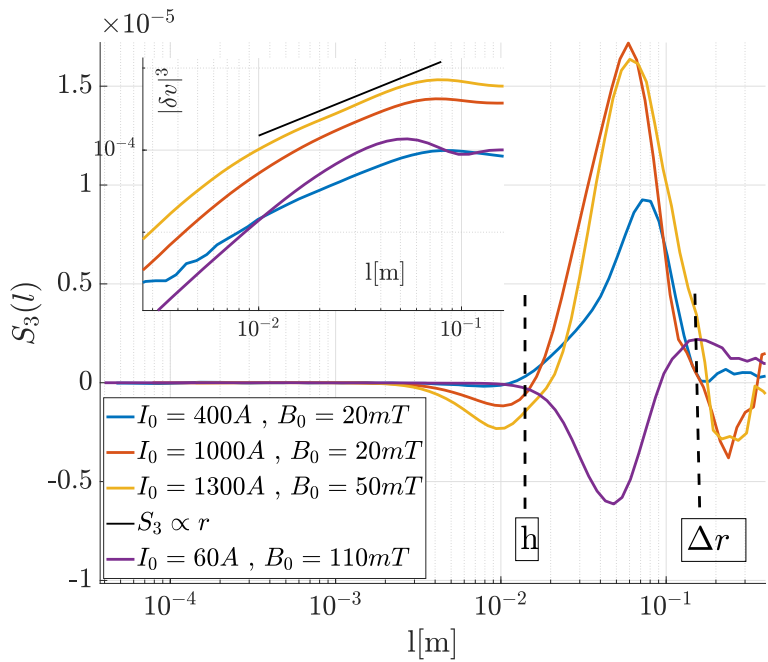}
    \caption{Structure function $S_3(l)$ at four different points in the parameter space. Inset: log-log plot of the modified structure functions $\langle |\delta V_L|^3\rangle$. $Ha\sim 0.57B_0$ with $B_0$ in $mT$.}
    \label{struc}
\end{figure}

This $k^{-5/3}$ spectrum could correspond either to a direct $3D$ turbulent cascade of energy as predicted by Kolmogorov, or to the Kraichnan prediction of an inverse cascade of energy towards the large scales. 

In order to access the energy flux characterizing the turbulent cascade, one can compute the third order structure functions of the velocity field defined by:
\begin{equation}
S_3(\ell)=\langle ({\bf u} ({\bf x, \ell+d\ell})-{\bf u({\bf x, \ell})})^3\rangle_x
\end{equation}
where $\langle ...\rangle_x$ denotes a spatial average. Since we do not have access to well resolved spatial measurements, we rather computed $S_3$ from very long time series of the velocity field acquired from the potential probes. To do so, we introduce 
 the velocity increment $\delta u(t,\tau) = {\bf u} (t+\tau)-{\bf u}(t)$ as the difference between the
values of the flow velocity at time  $t+\tau$ and $t$. One can then take advantage of the strong mean azimuthal flow to transform the quantity $S_3(\tau)=\langle\delta u(t,\tau)^3\rangle_t$ into $S_3({\bf \ell})$ by use of the  Taylor hypothesis $\ell=\tau\overline{u_\theta}$.
Depending on the direction of the flux of energy (in the $k$-space), two different behaviors are possible: in 3D turbulence, the famous Kolmogorov $4/5$ law $S_3(\ell)=-\frac{4}{5}\epsilon \ell$ is expected from the injection scale to the viscous dissipation scale. In the case of an inverse cascade, Kraichnan rather predicted an inverse cascade towards large scale with a positive structure function $S_3(\ell)=-\frac{3}{2}\epsilon \ell$ where $\epsilon$ is now negative.

Fig.\ref{struc} shows typical structure functions $S_3(l)$, obtained for the same set of parameters as for Fig.\ref{spectrum}. They all correspond to  different physical situations. For $B_0=20$mT and $I_0=400A$, the energy is injected at a scale slightly smaller than $h$. It then essentially cascades towards the large scales, as evidenced by the positive value of $S_3(\ell)$ and its linear dependance $S_3(\ell)\propto \ell$ in the range $\ell=[h-\Delta r]$. The corresponding $k^{-5/3}$ spectrum of the PSD therefore corresponds to an inverse cascade of energy as predicted by Kraichnan. \\

For larger forcings ($I_0B_0=20$ or $I_0B_0=65$), the structure functions now clearly exhibit negative values in the range $\ell=1mm - 30 mm$, and  positive values for $\ell>3cm$. The injection energy presumably takes place at an intermediate scale, leading to a (limited) scale separation  $\Delta r>\ell_{inj}> h$ which is typical of thin layer turbulence. The negative values of $S_3(\ell)$  are therefore the signature of a direct cascade of energy which takes place from $\ell_{inj}$ to smaller scale. On the other hand, some part of the energy also cascades upscale from $\ell_{inj}$ to $\Delta r$, as shown by the positive values of $S_3$ in this range. 
This split cascade of energy, in which the energy is injected at an intermediate scale $h<\ell_{inj}<\Delta r$ and cascades in both directions, is very similar to the one predicted for turbulence in very thin layers when the injection scale is much larger than the  thickness $h$ but much smaller than the largest scale (here $\Delta r$) \citep{Celtani10, Benavides17, Xia11}. Here, the natural integral scale for the $3D$ structures is the height of the disc, such that we expect $\ell_{inj}$ not too far from  $h$. In this regime, the $k^{-5/3}$ spectrum of the PSD for $k>2\pi h$ describes a forward cascade of energy, while the rest of the spectrum still corresponds to the inverse cascade. Note that in thin layer turbulence, the forward cascade of energy at small scale is accompanied by a co-directional cascade of enstrophy in the range $[l_{inj}-h]$. This enstrophy spectrum is not visible in our data, probably because of the very limited scale separation of the experiment.\\

Finally, the regime observed at $B_0=110mT, I_0=60A$ and reported in Fig.\ref{spectrum} and Fig.\ref{struc} obviously corresponds to a different regime than the double-cascade described above. These parameters are located deep inside the regime  of low frequency oscillations  described in section $4.5$.  When  this large scale instability occurs, it produces bi-dimensional vortices of the size of the gap $\ell\sim\Delta r$ which are advected by the mean flow and periodically passes through the probe. These vortices can be regarded as a type of condensate resulting from the inverse cascade \citep{Sommeria86}. The $k^{-3}$ spectrum shown in Fig.\ref{spectrum} may therefore be  a signature of this condensate, which takes the form of localized vortices traveling along the azimuthal direction.

 This scenario is very different from the one given by Messadek et al, where the  $k^{-3}$ spectrum is interpreted as an inverse cascade of energy modified by the strong Hartmann damping when the dissipation in the Hartmann layer is dominant at each wavenumber within the inertial range. 
 We however emphasize that because the low frequency oscillations strongly dominate the dynamics of the flow at large scale, the values of the structure function in the range $\ell=1cm - 10 cm$ probably does not accurately describe the energy flux. It is then difficult to determine which of these scenarios is correct. In any case, this regime is  associated with a very small energy flux at scales smaller than $h$.

\subsection{Transition from direct to  inverse cascade}

\begin{figure}
    \centering
    \includegraphics[scale=0.35]{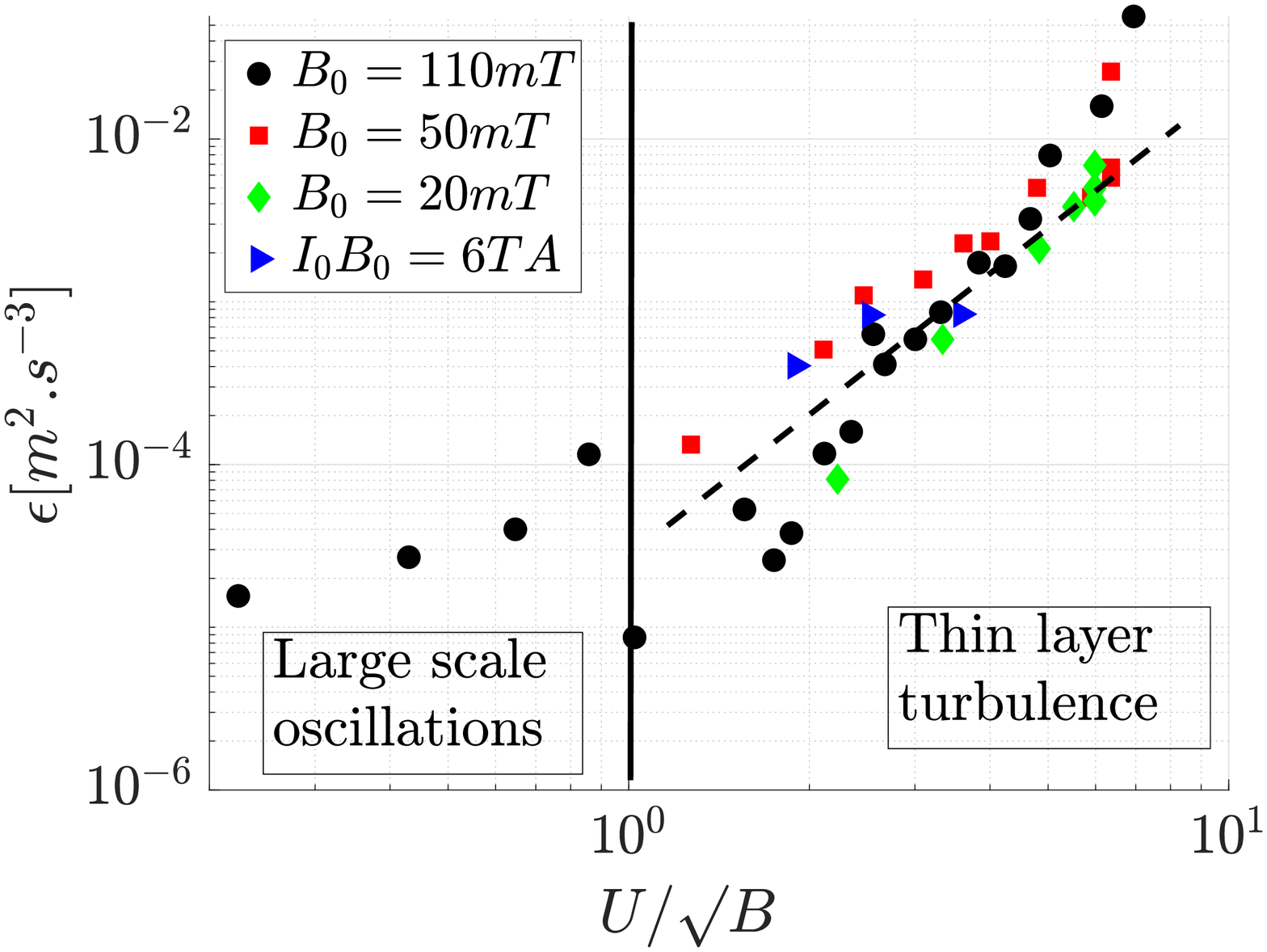}
    \includegraphics[scale=0.3]{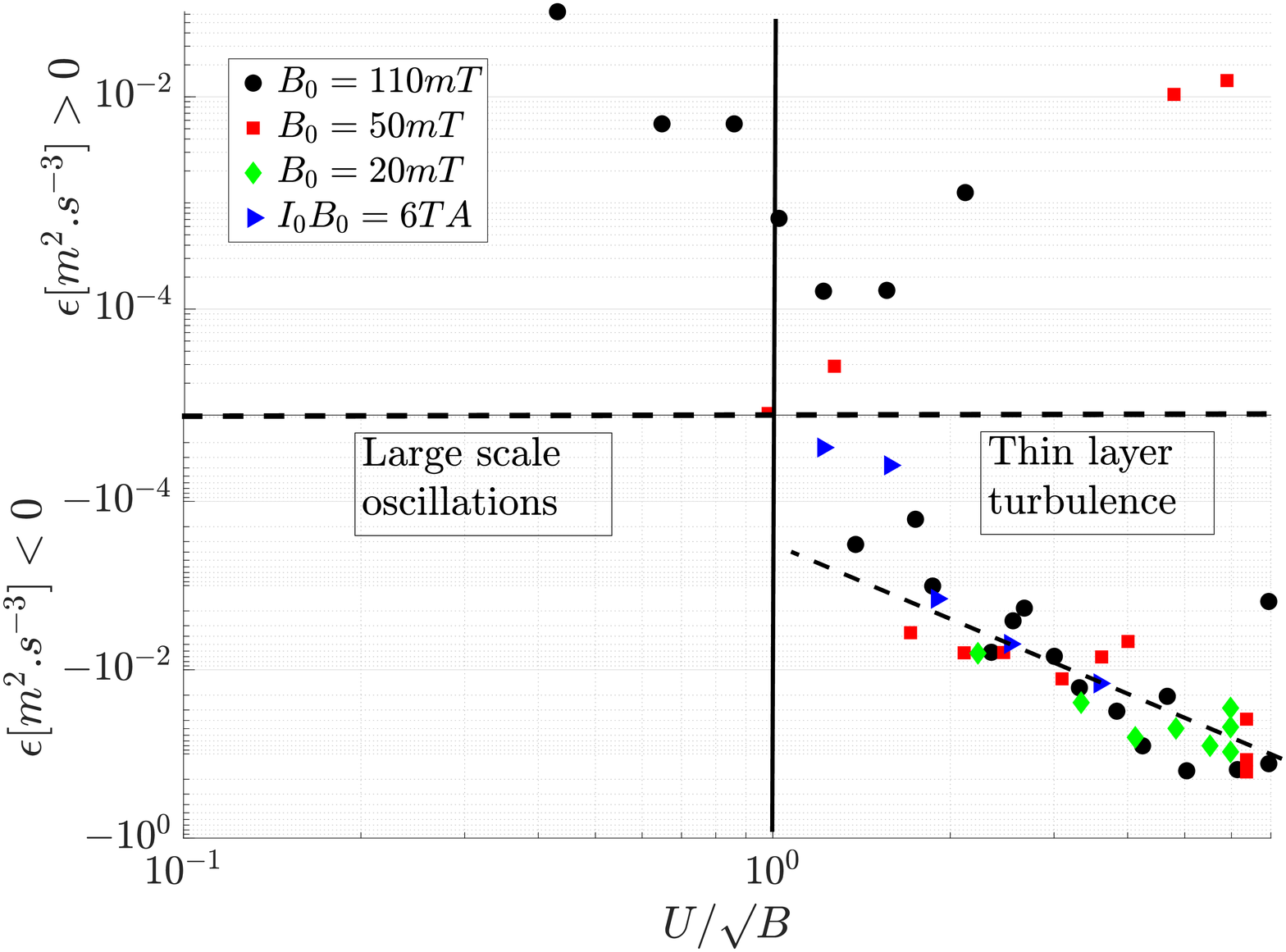}
    
    \caption{(a) Energy flux  at  large wavenumber  ($\ell=[10^{-3} - 10^{-2}]$m) as a function of $\overline{U_\theta}/\sqrt{B_0}$, showing a direct cascade of energy. (b) Energy flux  at  small wavenumber  ($\ell=[10^{-2} - 10^{-1}]$m) as a function of  $\overline{U_\theta}/\sqrt{B_0}$, showing a transition from a direct to an inverse cascade of energy. The two black dashed lines correspond to the scaling $|\epsilon|\propto \overline{U_\theta}^3$. $Ha\sim 0.57B_0$ with $B_0$ in $mT$.}
    \label{flux}
\end{figure}

In the recent years, the question of the transition from direct to inverse energy transfer in quasi-2D systems has been at the center of many studies, as summarized in the review of \citet{Alexakis18}.
To follow the transition between these different regimes, we estimate the energy flux at a given scale $\ell^*$ by computing the quantity:

\begin{equation}
\epsilon=-\langle \frac{S_3(\ell)}{\ell} \rangle
\end{equation}

\noindent where the average is taken on a given range of $\ell$ centered on $\ell^*$. With  this definition, a positive (resp. negative) value of $\epsilon$ corresponds to a flux of energy towards the small (resp. large) scales. Fig.\ref{flux}(a) shows the flux $\epsilon_f$ calculated on the range $\ell=10^{-3}-10^{-2}$m at small scale, while Fig.\ref{flux}(b) shows the flux $\epsilon_i$  calculated on the range $\ell=10^{-2}-10^{-1}$m, at large scale. We plot this energy flux as a function of the ratio $\overline{U_\theta}/\sqrt{B_0}$ which rescales on a single curve the data obtained for different Hartmann numbers.

Two different states are clearly identified: the regime generated at $\overline{U_\theta}/\sqrt{B_0}<1$ by the large scale oscillations of the flow is associated with a very weak flux of energy at small scale (Fig.\ref{flux}(a)) and a condensate at large scale ($S3>0$, see Fig.\ref{flux}(b)). For $\overline{U_\theta}/\sqrt{B_0}>1$, this regime is replaced by a state characteristic of thin layer turbulence: scales smaller than $h$ are characterized by a direct cascade of energy ($\epsilon>0$, see Fig.\ref{flux}(a)). The energy flux strongly increases with the velocity, and seems to follow a turbulent scaling $\epsilon\propto \overline{U_\theta}^3$. At large scales $h<\ell<\Delta r$, the flux is negative in this regime, as expected for an inverse cascade of energy and follows the same scaling $|\epsilon|\propto \overline{U_\theta}^3$ but with a larger amplitude: most of the injected energy cascades upscale.

At $\overline{U_\theta}/\sqrt{B_0}\approx 1$,  a  sharp bifurcation is observed between these two states. Although it describes a bifurcation between the condensate and some sort of thin layer turbulence, it is interesting to note that it takes the form of a second-order phase transition, similarly to what was observed in numerical shell models \citep{Benavides17} for the transition  between direct and inverse cascade.


\section{Concluding remarks}

The thin disk geometry $h/\Delta r\sim 0.1$ of the KEPLER experiment, combined with the use of a very large applied current and a large horizontal size  allows us to study regimes of electromagnetically-driven flow previously unexplored. Three different regimes were observed, depending solely on the values of the interaction parameter $N_t$ and the Reynolds number $Re$ compared to  their critical values $N_t^c\simeq10,Re^c\simeq9.10^4$.\\

At large interaction parameter ($N_t>N_t^c$) but small Reynolds number ($Re<Re^c$), the Lorentz force is balanced by viscosity, leading to the classical laminar solution, in which the velocity is independent of the magnetic field at the lowest order. In our thin disc geometry, this solution exhibits a marginally stable profile $\Omega\propto r^{-2}$. 

At smaller interaction parameter ($N_t<N_t^c$) and moderate Reynolds number ($Re<Re^c$), this laminar solution bifurcates to a regime in which the Lorentz force is now balanced by inertia, leading to a non-trivial dependence $u_\theta \propto (IB)^{2/3}r^{-1/3}f(\Delta r/r)$. Although dominated by inertia, the azimuthal flow is still laminar, and strongly influenced by the {\it laminar} Hartmann-B\"odewadt layers generated near the endcaps. The radial flow is non-negligible, especially near the boundaries, but the angular velocity  $\Omega\propto r^{-4/3}$ corresponds to {\it quasi-keplerian} rotation.

If the magnetic field remains moderate ($N_t<10$), a transition to turbulence occurs at large Reynolds number ($Re>Re^c$). This regime $u_\theta\propto\sqrt{IB/r}\log(Re)$ is now fully driven by the turbulent Reynolds stress and relies on the turbulence of the Hartmann-B\"odewadt boundary layers. For this reason, it may  be regarded as a regime of {\it strong turbulence}, somehow analogue to the {\it ultimate regime} observed in Taylor-Couette flows. It shows strong fluctuating velocity field and an efficient transport of angular momentum. The velocity profile exhibits a true {\it Keplerian rotation rate}, in which the angular velocity is $\Omega\propto r^{-3/2}$. This regime provides a very interesting analogue of the flow in accretion discs, which can be roughly described as Keplerian turbulent flows in thin magnetized discs.

Finally, in the small region of the parameter space defined by $N_t>10, Re>Re_c$, we report the existence of a large scale instability of the flow generating low-frequency perturbations at the largest scale $\ell_{inj}\approx 10 cm$. In this regime of bi-dimensional turbulence, third-order structure functions indicate that  the energy mostly condensate at the largest available scale, with only very weak energy flux at small scale and  fluctuations  mostly confined in the horizontal plane.

On the contrary, the {Keplerian turbulent regime} is characterized by a quasi-bidimensional flow, in which the energy is injected at an intermediate scale $h<\ell_{inj}< \Delta r$. Because of the {\it thin layer} geometry, this injected energy also cascades towards the  large horizontal scales. But in addition to this inverse cascade of energy, a direct forward cascade of energy towards scales smaller than $h$ is also observed, reproducing the phenomenology expected in thin layer turbulence. Surprisingly, the transition from one regime to the other occurs quite abruptly, leading to  a well defined critical threshold for the occurence of inverse (respectively direct) cascade of energy towards large (resp. small) scales.\\

From the astrophysical point of view, it is important to emphasize that the Kepler experiment \textit{is not} a laboratory study of the magneto-rotational instability. Indeed, MRI experiments aim at generating turbulence from an hydrodynamically stable pre-existing Keplerian flow, while the Keplerian turbulence in our system results from the instability of top and bottom boundary layers. This active role of the vertical velocity gradient shares some interesting similarities with the vertical shear instability considered for weakly ionized accretion disks \citep{Nelson13}. In consequence, the KEPLER experiment can rather be seen as an attempt to circumvent the current difficulties encountered by laboratory MRI experiments in generating magnetized Keplerian turbulence, in order to directly study the angular momentum transport or induction properties of accretion discs: the thin disc geometry together with the use of a moderate magnetic field provides an interesting analogue to these astrophysical objects. In addition, the angular momentum is not injected at the radial boundaries of the system, but re-inforced in volume by the Lorentz force. Compared to classical Taylor-Couette flows, we believe that studying the detailed angular momentum transport in such a configuration may offer new insights on this matter for turbulent astrophysical flows. More precisely, the fact that a \textit{ true} Keplerian rotation rate can be obtained in a turbulent regime is a significant advantage compared to relatively stable \textit{quasi-keplerian} Taylor-Couette flows.

Finally, an appealing continuation of this work would be to design a larger version of the experiment for which the magneto-rotational instability is possibly accessible. Several remarks can be made on the feasibility of this. First, as the electromagnetic forcing involves no moving parts, it reduces the risk of leaks compared to classical Taylor-Couette set-ups.
Using liquid Sodium ($\mu_0\sigma = 12 s.m^{-2}$), a larger horizontal size (L = 50cm), and a slightly larger electromagnetic forcing ($I_0 = 5000A$ and $B_0 = 0.15T$), equation (4.8) yields a typical velocity of $6m.s^{-1}$ leading to $Rm \sim 40$, which may be sufficient for the observation of MRI. Note however that in such configuration, the effect of the MRI would be to increase the angular momentum transport rather than triggering the turbulence. This means that the presence of turbulent fluctuations inherent to the generation of a Keplerian flow in our setup may very well increase the onset through turbulent diffusion, or complicate the measurements of angular momentum transport induced by MRI. Besides strong induction effects, the MRI also requires that the applied magnetic field is not too strong, namely $\bf{k}\cdot\bf{V_A}< \Omega$ with $V_A = B_0/\sqrt{\mu_0\rho}$, the Alfven velocity. As $\Omega$ decreases radially, this condition would be easily fulfilled, but only in the inner part of the flow. 
In this perspective, a numerical modeling of the experiment described here would be interesting, because it would clarify the critical onset for the occurrence of MRI, and help understanding the various regimes described in this paper. 
Rather than using liquid sodium, our experiment can also be half-filled with liquid Gallium, half-filled with Mercury, thus providing a quasi-free surface condition for the bottom Mercury (capillary length $l_c\sim 0.6mm$). At large currents, the azimuthal flow is expected to be an order of magnitude faster than the non-dispersive gravity waves generated at the Gallium-Mercury interface. This would provide an interesting shallow-water analogue of compressible turbulence in the presence of shock-waves, a situation once again encountered in accretion disks and in the interstellar medium.

\textbf{Acknowledgements.}  We thank A. Alexakis for useful discussions on the 2D turbulence, N. Garroum for his technical assistance, and our anonymous referees for their very constructive input.
\textbf{Funding.} This work was supported by funding from the French program 'JCJC' managed by Agence Nationale de la Recherche (Grant ANR 19-CE30-0025-01); the CEFLPRA contract 6104-1 ; and the Institut Universitaire de France.
\textbf{Declaration of interests.} The authors report no conflict of interest.
\textbf{Author ORCID.}M. Vernet, https://orcid.org/0000-0003-3128-0650; M. Pereira, https://orcid.org/0000-0001-9122-4082; S. Fauve, https://orcid.org/0000-0003-2379-6737; C. Gissinger, https://orcid.org/0000-0002-1734-6716.


\bibliographystyle{jfm}
\bibliography{Kepler}




\end{document}